\documentclass{emulateapj}

\shorttitle{Coherent structures in fast solar wind}
\shortauthors{Perrone et al.}

\usepackage{color}
\begin{document}

\title{Coherent structures at ion scales in fast solar wind: {\it Cluster} observations}

\author{D. Perrone$^{1}$, O. Alexandrova$^2$, O. W. Roberts$^3$, S. Lion$^2$, C. Lacombe$^2$, A. Walsh$^1$, M. Maksimovic$^2$, I. Zouganelis$^1$}
\affil{$^1$ European Space Agency, ESAC, Madrid, Spain. \\
$^2$ LESIA, Observatoire de Paris, PSL Research University, CNRS, Sorbonne Universités, UPMC Univ. Paris 06, Univ. Paris Diderot, Sorbonne Paris Cit\'e, France. \\
$^3$ European Space Agency, ESTEC, Noordwijk, Netherlands.}

\begin{abstract}
We investigate the nature of magnetic turbulent fluctuations, around ion characteristic scales, in a fast solar wind stream, by using {\it Cluster} data. Contrarily to slow solar wind, where both Alfv\'enic ($\delta b_{\perp} \gg \delta b_{\parallel}$) and compressive ($\delta b_{\parallel} \gg \delta b_{\perp}$) coherent structures are observed \cite[]{per16}, the turbulent cascade of fast solar wind is dominated by Alfv\'enic structures, namely Alfv\'en vortices, with small and/or finite compressive part, with the presence also of several current sheets aligned with the local magnetic field. Several examples of vortex chains are also recognized. Although an increase of magnetic compressibility around ion scales is observed also for fast solar wind, no strongly compressive structures are found, meaning that the nature of the slow and fast winds is intrinsically different. Multi-spacecraft analysis applied to this interval of fast wind indicate that the coherent structures are almost convected by the flow and aligned with the local magnetic field, i.e. their normal is perpendicular to {\bf B}, that is consistent with a two dimensional turbulence picture. Understanding intermittency and the related generation of coherent structures could provide a key insight into the nonlinear energy transfer and dissipation processes in magnetized and collisionless plasmas. 
\end{abstract}

\section{Introduction}

Turbulence is a complex, fascinating and highly non-linear process ubiquitous in nature. The dynamical evolution of a turbulent system is a consequence of the nonlinearity responsible for the coupling of many degrees of freedom, which leads the system far from thermal equilibrium. In typical hydrodynamic systems, the energy, injected at large scales, is transferred self-consistently towards smaller scales, where it finally can be dissipated \cite[]{kol41}. Moreover, the turbulent activity becomes more and more inhomogeneous and non-uniform as the energy arrives to smaller and smaller scales. This phenomenon, named intermittency \cite[]{fri95}, is due to the presence of coherent structures, i.e. filaments of the vorticity field localized in space but covering scales from about the integral scale to the dissipation scale and with a characteristic tube-like structure \cite[]{she90,fri95}. These intermittent events contain most of the energy of the flow. Although for classical fluids the turbulence is fairly well understood, in plasma physics this process represents one of the most spectacular and unsolved problems, where both cross-scale couplings and kinetic effects are present. In this case, indeed, the energy, injected at large scales, progressively decays towards smaller scales, where kinetic effects (heating, particle acceleration and so on) dominate the plasma dynamics.

Thanks to the support of many space missions, we have the unique opportunity to analyze directly the dynamics of a natural turbulent plasma: the solar wind, a continuous, but highly variable, weakly collisional and multi-component plasma outflow from the Sun that travels at high speed. `In situ' measurements generally show that the interplanetary medium is in a state of fully developed turbulence, characterized by a multi-scale nonlinear behavior \cite[]{bru13}. In the inertial range the turbulent magnetic field cascade manifests a power law similar to the fluid behavior \cite[]{kol41}. However, around ion characteristic scales, a change in the spectral shape is observed, with the presence of a steeper spectrum \cite[]{lea98,lea00,bal05,smi06,ale07,ale08,bou12,bru14,lio16} and an increase of the magnetic compressibility \cite[]{ale07,ale08,ham08,sal12,kiy13,tel15,lac17}. In this frequency range, often called dissipation range, the plasma dynamics is governed by kinetic effects, namely strong anisotropies in the ion velocity distributions, with preferential perpendicular heating and parallel accelerated particles with respect to the background magnetic field. At scales smaller than the ion characteristic scales and up to a fraction of the characteristic electron lengths, another general spectrum is observed, whose interpretation is still controversial \cite[]{ale09,ale12,sah10,sah13}.

A very important aspect of the solar wind turbulent cascade is intermittency, due to the non-Gaussian and bursty nature of the turbulent fluctuations, with the non-Gaussianity that increases towards smaller scales. Therefore, as in the fluid case, also in the solar wind the energy is not uniformly distributed in space \cite[]{bru01}, but is localized in coherent structures, i.e. structures characterized by a phase synchronization among a certain number of scales. A clear link exists between intermittency, non-Gaussianity and phase coherence. \cite{kog07} have shown, in the solar wind turbulence near the Earth's bow shock, that exists a similar behavior between the phase coherence index and the kurtosis (flatness), reflecting a departure from Gaussianity of the probability density function of the magnetic field fluctuations, where the non-Gaussianity of the fluctuations is a clear signature of intermittency. 

During the last decades, the presence of planar structures, such as current sheets, rotational discontinuities and shocks \cite[]{vel99,vel99b,sal09,gre12,per12,gre14}, was considered to be the principal cause of intermittency in solar wind at ion scales. Recent studies have shown that other types of coherent structures also contribute to the intermittency phenomenon in the solar wind turbulence. \cite{lio16} have shown the presence of Alfv\'en vortex-like structures in a fast solar wind stream by using \textit{Wind} measurements. Moreover,  a study by \cite{rob16} using multi-satellite measurement from \textit{Cluster} spacecraft has shown a well-defined Alfv\'en vortex in a slow solar wind stream. These structures occur close to ion characteristic scales, similar to what happens to the vortices observed in the Earth's and Saturn's magnetosheaths \cite[]{ale06,ale08b}. 

More recently, a statistical analysis of coherent structures around ion scales in a slow solar wind stream has been performed by \cite{per16}, using \textit{Cluster} measurements. This study has shown, for the first time, that different families of coherent structures participate to the intermittency at ion scales in slow solar wind, such as compressive structures, i.e. magnetic holes, solitons and shocks; and alfv\'enic structures in form of current sheets and vortices. These last ones can have an important compressive part and they are the most frequently observed during the analyzed interval. All the observed structures are field aligned with normals perpendicular to the local mean magnetic field, that is consistent with two dimensional ($k_\perp \gg k_\parallel$) turbulence. Moreover, although most of the structures are merely advected by the wind, the 25\% of the analyzed structures propagate in the plasma reference frame. 

Despite the fact that several studies have been performed to understand the complex behavior of the solar wind, the nature of the turbulent fluctuations around ion scales and the dissipation in such collisionless medium still remain an open question.  The purpose of the present paper is to shed light on the nature of the turbulent fluctuations around ion scales in fast solar wind by using multi-point measurements from \textit{Cluster} spacecraft. The fast solar wind is generally characterized by a higher proton temperature and a lower density with respect to the slow solar wind. Other differences between the two streams are the composition, the Alfv\'enic content and the anisotropies in ion and electron temperatures.

In the present work, first we focus on the turbulent character of the considered stream and on the phase coherence between the components of the magnetic field, by using wavelet analysis \cite[]{far92,tor98}. Wavelet transforms are a mathematical method, which allow unfolding a signal, or a field, into both time and scale at once. The wavelet analysis can be performed locally on the signal, as opposed to the Fourier transform, which is inherently nonlocal. By expanding the signal in a set of functions that are localized in time as well as in frequency, it is possible to highlight the presence of regions characterized by intermittency in the considered stream, thus studying the `texture' of the turbulence. Then, we investigate in detail the magnetic field fluctuations close to the ion scales by using the {\it timing} method for the analysis of multi-satellite data \cite[]{sch98,per16}. The considered interval of solar wind appears to be characterized by the presence of coherent structures. Moreover, by applying the {\it multi-point signal resonator} technique \cite[]{nar11_res,nar11a} to the same magnetic fluctuations, we verify the applicability of the {\it k-filtering} in the case of strong turbulence, i.e. in the presence of coherent structures.

Finally, as a result of the statistical studies on the coherent structures observed in the stream, we find that the ion scales are dominated by Alfv\'en vortices, with small and/or finite compressive part. Moreover, we observe the presence also of several current sheets aligned with the magnetic field, almost convected by the wind. The comparison of these results with the results presented in \cite{per16} suggests that the physics that governs the ion scales of fast and slow solar wind is quite different. 

The paper is organized as follows: in Section~\ref{sec:data} we describe the selected data interval of fast solar wind in terms of plasma parameters and turbulent behavior; and in Section~\ref{sec:intermittency} we discuss the concepts of intermittency and phase coherence. In Section~\ref{sec:structures} we present some examples of detected coherent structures and theoretical models in order to explain the observations. In Section~\ref{sec:analysis} we determine spatial orientation and plasma frame velocities of the observed structures by using multi-satellite analysis; and, in Section~\ref{sec:conclusion} we summarize the results and we discuss our conclusions. Finally, in the Appendix, we present the results of the {\it multi-point signal resonator} technique.

\section{Fast Solar Wind Interval}
\label{sec:data}

We consider an interval of about 40 minutes (17:30-18:10~UT) of undisturbed solar wind from {\it Cluster} spacecraft on 2004 January 31st. It is a stream of fast solar wind, with a mean speed of about 600~km/s, characterized by a mean magnetic field of about 8.3~nT, a mean proton density of about 3-4~cm$^{-3}$, and a mean proton temperature of about $\sim 37$~eV, with $T_{\|,p} \sim 30$~eV and $T_{\perp,p} \sim 41$~eV. In terms of characteristic plasma scales, the proton Larmor radius, defined as the ratio between the perpendicular proton thermal speed and the proton cyclotron frequency, is $\rho_{p} \sim$ 109~km; while the proton inertial length, defined as the ratio between the light speed and the proton plasma frequency, is $\lambda_p \sim$ 121~km. Finally, the proton plasma beta, $\beta_p$, defined as the ratio between proton kinetic pressure and magnetic pressure, has an averaged value in the interval of about $0.8$, with several regions where $\beta_p > 1$.

Although some caveats exist for the electron moments by using the Plasma Electron and Current Experiment (PEACE) \cite[]{joh97,faz09} in this interval of fast solar wind, the mean parallel and perpendicular temperatures are about $T_{\|,e} \sim 18$~eV and $T_{\perp,e} \sim14$~eV, respectively. Therefore, the electron Larmor radius is $\rho_e \sim 1.5$~km. By using the Waves of High Frequency and Sounder for Probing of the Electron Density by Relaxation (WHISPER) experiment \cite[]{dec01}, the electron density is known ($\sim$ 4~cm$^{-3}$) with a resolution of 1.5~s and the electron inertial length can be derived ($\lambda_e \sim$ 2.6~km).

\begin{figure}
\begin{center}
\includegraphics [width=7.5cm]{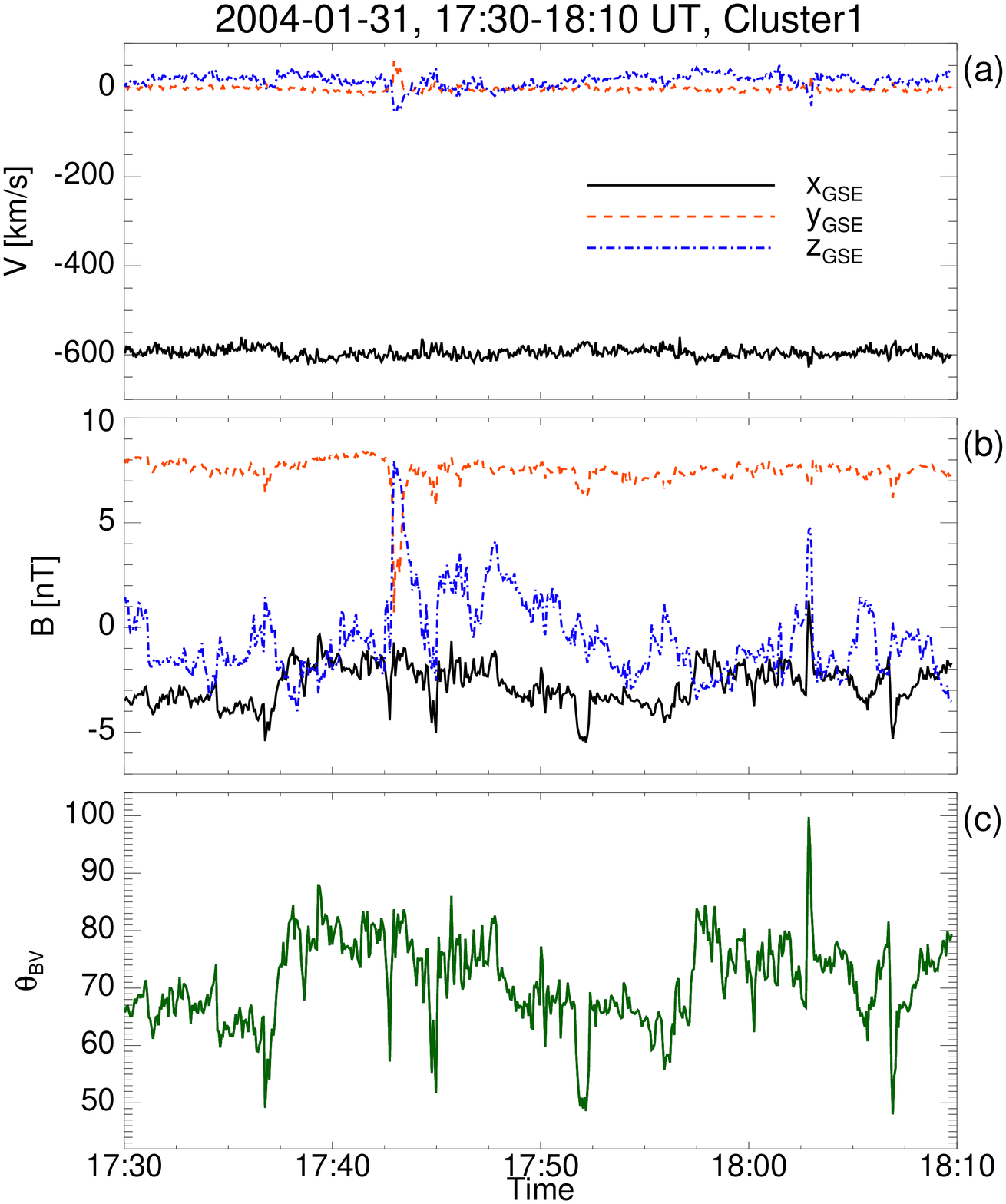}
\caption{Time interval of fast solar wind as measured by the \textit{Cluster1} satellite on January $31^{st}$, 2004 (17:30-18:10~UT). Panel (a) and (b) show the velocity, $V$, and magnetic, $B$, field components in GSE, respectively; while panel (c) displays $\theta_{BV}$, the angle between the two previous fields.}
\label{fg:stream}
\end{center}
\end{figure}

Figure~\ref{fg:stream} gives a brief overview of the considered interval from C1. In particular, in panel (a) we show the three components of the velocity field from the Hot Ion Analyser (HIA) sensor of the Cluster Ion Spectrometry (CIS) with a resolution of 4 seconds \cite[]{rem01}. Here, $x$ (black), $y$ (red) and $z$ (blue) denote the Geocentric Solar Ecliptic (GSE) coordinate system, i.e. $x$ component points towards the Sun and $z$ axis is perpendicular to the plane of the Earth's orbit around the Sun (positive North). The $v_y$, that is the component in the direction anti-parallel to the direction of Earth's motion, has been corrected for the $\sim 30$~km/s aberration produced by the orbital speed of the spacecraft and Earth around the Sun. Moreover, panel (b) displays the raw data of the magnetic field vector, from the Fluxgate Magnetometer (FGM) \cite[]{bal01}, where the three components are also given in GSE, by using the same colors of panel (a). Although in this case the data are represented with the same resolution of the velocity field in panel (a), i.e with a 4 second resolution, in the following part of the paper we will use the highest sampling time of the FGM instrument in nominal mode (22~Hz) to properly describe ion scales. Finally, panel (c) of Figure~\ref{fg:stream} shows the temporal evolution of the angle between magnetic and velocity fields, $\theta_{BV}$. The large values of the angle, $\theta_{BV} > 50^{\circ}$, indicate that the two vectors are approximately perpendicular, suggesting that there is no connection of the analyzed stream of solar wind with the Earth's foreshock: indeed, the electrostatic waves, typical of a magnetic connection, are not observed on WHISPER during this interval (not shown here). 

\begin{figure}
\begin{center}
\includegraphics [width=7.8cm]{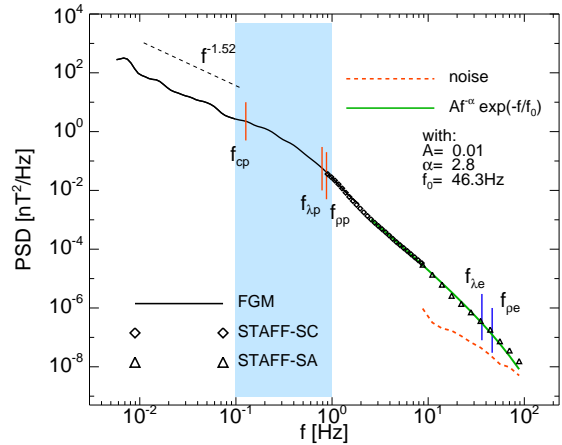}
\caption{Power spectral density of total magnetic fluctuations measured on C1, by FGM (up to $\sim 1$~Hz, solid line), STAFF-SC (up to $\sim 9$~Hz, diamonds) and STAFF-SA ($f \ge 8$~Hz, triangles). The red dashed line shows the STAFF-SA noise level on C1. The power-law fit in the MHD range is displayed by the black dashed line, while the spectrum at scales smaller than the ion characteristic scales is described by $Af^{-\alpha}\exp(-f/f_0)$ (green solid line). 
Vertical solid bars indicate the proton (in red) and the electron (in blue) characteristic scales (assuming $f_{\rho_e}=f_0$), while the blue filled band denotes the range of scales, $f \in [0.1,1]$~Hz, around ion scales.}
\label{fg:total_spectrum}
\end{center}
\end{figure}

In order to quantify the turbulence we compute the power spectrum of the magnetic fluctuations, up to electron scales. Figure~\ref{fg:total_spectrum} shows the total power spectral density (PSD), $S(f)=\sum_{i=x,y,z} S_i(f)$, where $S_i$ is the PSD of each component of the magnetic field and is defined as
\begin{equation}
S_i(\tau) = \frac{2 \delta t}{N} \sum_{j=0}^{N-1} | {\mathcal W}_i(\tau,t_j) |^2 \ , \ \ \ i=x,y,z
\end{equation} 
begin $\delta t$ the time spacing and ${\mathcal W}_i$ the Morlet wavelet coefficients for different time scales $\tau$ and time $t_j$ \cite[]{tor98}
\begin{equation}
{\mathcal W}_i(\tau,t) = \sum_{j=0}^{N-1} B_i(t_j)\psi^{*}[(t_j-t)/\tau] \ .
\end{equation}
The frequency dependence is easily obtained using the $f=1/\tau$ relationship. 

We use the onboard FGM measurements up to $\sim 1$~Hz (solid line) and the Spatio Temporal Analysis of Field Fluctuation experiment/Search Coil (STAFF-SC) \cite[]{cor03}, with a resolution of 25~Hz in the frequency domain [0.35--9]~Hz (diamonds). Finally, we complete the analysis with the Spatio Temporal Analysis of Field Fluctuation experiment/Spectrum Analyser (STAFF-SA) on C1, which provides 4 seconds averages of the power spectral density of the magnetic fluctuations at 27 logarithmically spaced frequencies, between 8~Hz and 4~kHz (triangles). The red dashed line shows the instrumental STAFF-SA noise level that becomes significant for $f >88$~Hz, where the signal-to-noise ratio (S/N) is lower than 3. This region is omitted in Figure~\ref{fg:total_spectrum} to avoid any misunderstanding. 

The spectrum in Figure~\ref{fg:total_spectrum} shows the characteristic behaviour of the solar wind turbulent cascade \cite[]{bru14}. At low frequencies, in the MHD range, the distinctive behaviour of a power law, $\propto f^{-1.52}$, is observed (black dashed line). Then, around $0.3$~Hz, that is in-between the characteristic proton frequencies (i.e., cyclotron frequency, $f_{cp}$, and Doppler shifted proton Larmor radius, $f_{\rho_p}=v_{sw}/2\pi {\rho_p}$, and proton inertia length, $f_{\lambda_p}=v_{sw}/2\pi {\lambda_p}$) estimated under the assumption of wave-vector parallel to the plasma flow (vertical red bars), a change in the spectral shape is observed. At higher frequencies ($f > 0.3$~Hz), the spectrum is steeper and is well described by the exponential model proposed by \cite{ale12} for a general description of the whole turbulent spectrum at kinetic scales. The present model 
\begin{equation}
PSD(f) = A f^{-\alpha} \exp(-f/f_0)
\end{equation}
is composed by an exponential with a characteristic frequency $f_0$ and with a power-law pre-factor. Therefore, the three free parameters are (i) the amplitude $A$, (ii) the spectral index $\alpha$ and (iii) the cutoff frequency ($f_0$). The exponential model fitting is shown in Figure~\ref{fg:total_spectrum} by the green solid line, including also the parameters of the fit: $A \simeq 0.01$, $\alpha \simeq 2.8$ and $f_0 \simeq 46.3$~Hz. By considering $f_0$ related to the Doppler shifted electron Larmor radius, $f_{\rho_e}=v_{sw}/2\pi {\rho_e}$, in agreement with the general spectrum proposed by \cite{ale12}, $f_0 \in [0.74 f_{\rho_e}, f_{\rho_e}]$, we have $\rho_e \in [1.5, 2.1]$~km. This result is in agreement with $\rho_e$ estimated by using directly the electron perpendicular temperature.  

In Figure~\ref{fg:total_spectrum}, the blue filled band denotes the range of scales, $f \in [0.1,1]$~Hz, that are of interest for the investigation of the nature of the turbulent fluctuations around ion scales, which is the aim of the present work. Therefore, in the following part of the paper, we will focus on the high-resolution magnetic field data given by FGM on C1. 

\begin{figure}
\begin{center}
\includegraphics [width=7.8cm]{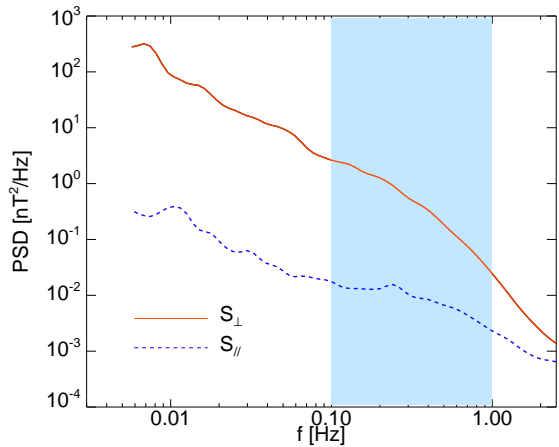}
\caption{Power spectral density of the perpendicular (red solid line) and parallel (blue dashed line) magnetic field fluctuations measured by FGM. The blue filled band denotes the range of scales $f \in [0.1,1]$~Hz, as in Figure~\ref{fg:total_spectrum}.}
\label{fg:anis_spectrum}
\end{center}
\end{figure}

Figure~\ref{fg:anis_spectrum} shows the PSD of the perpendicular (red solid line) and parallel (blue dashed line) magnetic field fluctuations. The variations of the magnetic field magnitude can be used as a proxy of the parallel (compressive) fluctuations ($\delta B_{\parallel}=\delta |B|^2/2B_0$, where $B_0$ is the mean magnetic field on the whole interval \cite[]{per16}), so the corresponding energy is 
\begin{equation}
{\mathcal W}_{\|}^2(\tau,t) = {\mathcal W}_{|B|}^2(\tau,t) \ , 
\end{equation}  
while the energy of the perpendicular (Alfv\'enic) fluctuations is defined as
\begin{equation}
\label{eq:W2-perp}
{\mathcal W}_{\perp}^2(\tau,t) = {\mathcal W}_{{\bf B}}^2(\tau,t) -  {\mathcal W}_{\|}^2(\tau,t) \ ,  
\end{equation}  
where ${\mathcal W}_{{\bf B}}^2(\tau,t)$ is the total energy of the magnetic fluctuations. The bias in the spectrum due to the quantization noise, not shown in the Figure, is $q^2/12 f_s \sim 4 \cdot 10^{-7}$~nT${^2}/$Hz \cite[]{wid08}, being $q$ the digitization of the instrument ($10^{-2}$~nT for FGM in the solar wind mode), $f_s$ the sampling frequency (22~Hz) and by assuming that the noise is uniformly distributed over the entire spectral range. However, although the quantization noise is very low, the comparison between FGM and STAFF data shows that the two spectra start to deviate for $f> 1-2$~Hz, meaning that the instrumental noise could become important at frequencies higher than $1-2$~Hz.  

Figure~\ref{fg:anis_spectrum} shows that, although in the inertial range the energy stored in the perpendicular direction is much stronger than the compressive one, around ion scales the compressive energy increases, meaning that the contribution of the parallel magnetic fluctuations becomes important at kinetic scales \cite[]{ale07,ale08,sal12,kiy13,per16} and, in particular, in the frequency range around the spectral transition ($f \in [0.1,1]$~Hz, blue filled band). The small bump in the compressive energy around 0.25 Hz corresponds to the satellite spin ($\tau =4$~s).

\section{Intermittency, non-Gaussianity and Phase Coherence}
\label{sec:intermittency}

\begin{figure}
\begin{center}
\includegraphics [width=8.8cm]{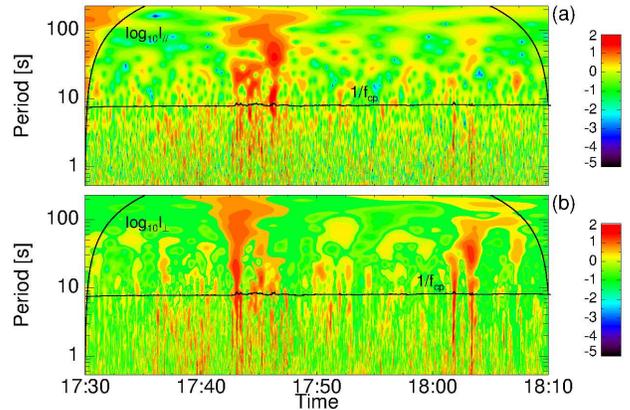}
\caption{Logarithmic contour plots of LIM, $I(\tau,t)$, for parallel (a) and perpendicular (b) magnetic field fluctuations. Horizontal black lines denote the proton cyclotron period, while the curved black lines indicate the cone of influence for the Morlet transform.}
\label{fg:lim}
\end{center}
\end{figure}

The spectra, in general, provide information about the global properties of the turbulent activity, but only a local analysis of the fluctuations enables to understand the details of turbulence. In this context, panels (a) and (b) of Figures~\ref{fg:lim} show the evolution of the Local Intermittence Measure (LIM) for the parallel and perpendicular magnetic energy, respectively. The LIM is defined as the energy of magnetic fluctuations, as a function of time and scales, normalized at each time point by a mean spectrum over the whole time interval:
\begin{equation}
I_{\|,\perp}(\tau,t) = \frac{| {\mathcal W_{\|,\perp}}(\tau,t) |^2}{\langle| {\mathcal W_{\|,\perp}}(\tau,t) |^2 \rangle_t} \ .
\end{equation}
The curved black lines, on each side of the plots in Figures~\ref{fg:lim}, represent the cone of influence where the Morlet coefficients are affected by edge effects \cite[]{tor98}.
Non uniform distribution of energy is observed in both parallel and perpendicular components with the appearance of localized energetic events covering a certain range of scales, which are easily recognized by red color. This is an indication of the presence of coherent structures in the system, that will be described in detail in Section~\ref{sec:structures}. These intermittent events are strictly connected with the strong variations of the magnetic field components (see for example the variation between 17:40 and 17:50 in panel (b) of Figure~\ref{fg:stream}), that is highlighted by the variation of the inverse of the proton cyclotron frequency (horizontal black lines in Figure~\ref{fg:lim}). 

\begin{figure}
\begin{center}
\includegraphics [width=8.8cm]{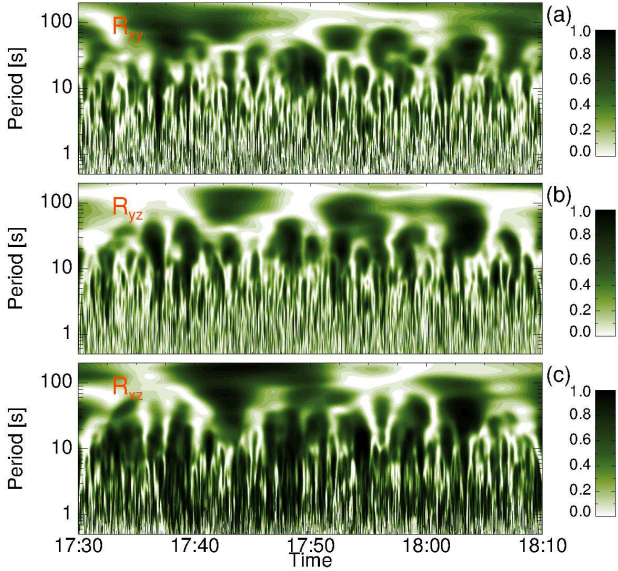}
\caption{Logarithmic contour plots of phase coherence, $R_{ij}(\tau,t)$, for $B_x$ and $B_y$ (a); $B_y$ and $B_z$ (b); and, $B_x$ and $B_z$ (c).}
\label{fg:cohe}
\end{center}
\end{figure}

Let's consider the phase coupling between the magnetic components during the analyzed time interval. Figure~\ref{fg:cohe} shows the phase coherence, $R_{ij}(\tau,t)$, between two components, $B_i$ and $B_j$, defined as \cite[]{gri04,lio16} 
\begin{equation}
R_{ij}^2(\tau,t) = \frac{\left| \mathcal S \left(\tau \mathcal W_i(\tau,t) \mathcal W_j^*(\tau,t) \right)\right|^2 }{\mathcal S \left(\tau \left| \mathcal W_i(\tau,t) \right|^2 \right) \cdot \mathcal S \left(\tau \left| \mathcal W_j^*(\tau,t) \right|^2\right)} \ ,
\end{equation}
where $\mathcal S$ is a compound smoothing operator for frequencies and time, $\mathcal S(\mathcal W(\tau,t))=\mathcal S_{\tau}(\mathcal S_t(\mathcal W(\tau,t)))$, with 
\begin{equation}
\mathcal S_{\tau}(\mathcal W(\tau,t)=\mathcal W(\tau,t)c_1^{-t^2\tau^2/2}
\end{equation} 
\begin{equation}
\mathcal S_{t}(\mathcal W(\tau,t)=\mathcal W(\tau,t)c_2 \Pi({0.6}/{\tau})
\end{equation} 
being $c_1$ and $c_2$ the normalization constants \cite[]{gri04}, $\Pi$ the rectangular function and $0.6$ the scale decorrelation length for the Morlet wavelet \cite[]{tor98}. By definition, the values of $R_{ij}(\tau,t)$ are between 0 (no coherence, in white) and 1 (full coherence, in black). 

We consider the phase coherence between magnetic components in a reference frame where $z$ is aligned with a magnetic field ${\bf B_0}$ averaged on the whole time interval (${\bf e_z} = {\bf e_b}$); $x$ is perpendicular to ${\bf B_0}$ in the  plane spanned by it and the radial direction (${\bf e_x} = {\bf e_b}\times{\bf e_r}$); and $y$ closes the right-hand reference frame (${\bf e_y} = {\bf e_b}\times{\bf e_x}$). In this case, we are considering the global frame defined on 40 minutes. In Sections \ref{sec:structures} and \ref{sec:analysis} a local frame will be assumed by considering a magnetic field at scales on the same order of scales of the individual turbulent structures. It is worth pointing out that the choice of a particular mean magnetic field could lead to significant differences in physical results \cite[]{che11,mat12,ten12}. However, in our case, both the magnetic frames produce almost the same results since the angle between the global and the local magnetic field is small (its histogram, not shown here, is peaked around 10$^{\circ}$). 

Figure~\ref{fg:cohe} shows the phase coherence of each couple of magnetic field components, where localized regions in time of high coherence are found, that cover a certain range of scales, including the frequency range $f \in [0.1,1]$~Hz ($\tau \in [1,10]$~s), as in the case of intermittency for both perpendicular and parallel magnetic energy (Figures~\ref{fg:lim}).

\begin{figure}
\begin{center}
\includegraphics [width=8.8cm]{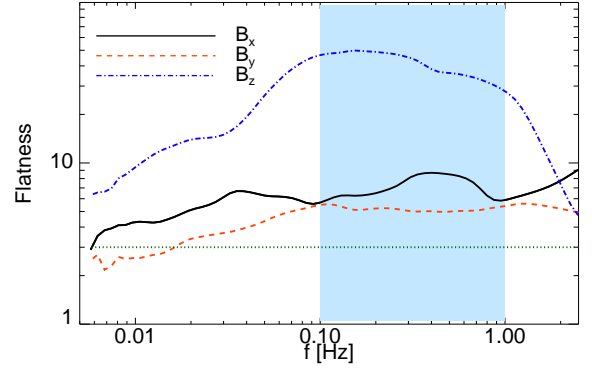}
\caption{Flatness, $\mathcal F_{i}(\tau)$, for $B_x$ (black solid line), $B_y$ (red dashed line) and $B_z$ (blue dot-dashed line). The horizontal green dotted line indicates the value of the flatness for a standard normal distribution, while the blue filled band denotes the range of scales $f \in [0.1,1]$~Hz.}
\label{fg:flatness}
\end{center}
\end{figure}

Keeping the same reference frame, we investigate the Gaussianity of the magnetic fluctuations as a function of scale (or frequency) by using the fourth-order moment of each component. We define the flatness (or kurtosis) of $B_i$ as
\begin{equation}
\label{eq:flat}
\mathcal F_{i}(\tau) = \frac{\langle {\tilde{\mathcal W_{i}}}(\tau,t)^4 \rangle}{\langle {\tilde{\mathcal W_{i}}}(\tau,t)^2 \rangle^2} \ , 
\end{equation}
where $\tilde{\mathcal W}$ is the real part of the wavelet coefficient and $\tau=1/f$. If $\mathcal F_{i}(\tau)=3$, the probability distribution function (PDF) of the corresponding component of magnetic field fluctuations is a standard normal distribution, while if $\mathcal F_{i}(\tau) >3$ the PDF is not a Gaussian distribution, showing fat tails. 

Figure~\ref{fg:flatness} shows $\mathcal F_{i}(f)$ for $B_x$ (black solid line), $B_y$ (red dashed line) and $B_z$ (blue dot-dashed line). The value of the flatness for a standard normal distribution (horizontal green dotted line) is given as reference. We observe that the flatness of both parallel and perpendicular magnetic field fluctuations departs from the normal distribution value, reflecting a non-homogeneous distribution of the turbulent fluctuations as already observed in the maps of the LIM (Figure~\ref{fg:lim}). 
Moreover, after an initial increase of $\mathcal F_{i}$, the flatness of both perpendicular ($\mathcal F_{x}$ and $\mathcal F_{y}$) and parallel ($\mathcal F_{z}$) fluctuations becomes nearly constant around ion scales (blue filled band). However, for $f \sim1$~Hz, i.e where the FGM data start to deviate from STAFF data, $\mathcal F_{z}$ becomes to decrease. This behaviour could be due to the fact that the noise becomes to be important for $f>1-2$~Hz. Indeed, the noise is expected to have Gaussian statistics; thus, $\mathcal F_{z}$ might approach the constant value expected for a Gaussian distribution. However, a decrease in flatness, related to the frequency location of the break, has been already observed in literature by \cite{wu13} and \cite{tel15}. In particular, \cite{wu13} observed a flatness decrease in all the magnetic components, arguing that it is of physical origin, due to an additional ingredient of incoherent dynamics.

Therefore, in addition to the intermittency and phase coherence, the ion scales appear also characterized by the departure from Gaussianity of the PDFs. However, the expected behavior for the intermittency to increase decreasing the scales is not observed in these particular range of scales, where the flatness is almost constant. Moreover, it is worth pointing out that $\mathcal F_{z}$ reaches a higher value of saturation with respect to $\mathcal F_{x}$ and $\mathcal F_{y}$ (even though most of the energy is stored in the $B_x$ component), meaning that the parallel direction could represent a preferential channel for the evolution of nonlinear effects at kinetic scales. 

To focus on the range of scale around ion scales and to compare the present analysis with the results in slow solar wind described in \cite{per16}, we use a bandpass filter based on the wavelet transform \cite[]{tor98,he12,rob13,per16} for the range $f \in[0.1,2]$~Hz, defined as
\begin{equation}
\label{eq:db}
\delta b_i(t) = \frac{\delta j \delta t^{1/2}}{C_{\delta} \psi_0(0)} 
\sum_{j=j_1}^{j_2} \frac{\tilde{\mathcal W_i}(\tau_j,t) }{\tau_j^{1/2}} \ , 
\end{equation}
where $j$ is the scale index and $\delta j$ is the constant step in scales; the factor $\psi_0(0) =\pi^{1/4}$ and the value of the constant $C_{\delta}$, derived from the reconstruction of a $\delta$ function using the Morlet wavelet, is 0.776 \cite[]{tor98}. Finally, $\tau(j_1) = 0.5$~s and $\tau(j_2) = 10$~s. 

\begin{figure*}
\begin{center}
\includegraphics [width=14.8cm]{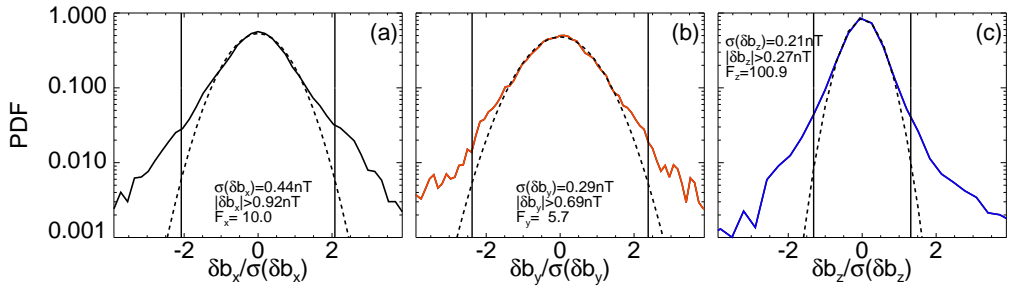}
\caption{PDFs of $\delta b_x$ (a), $\delta b_y$ (b) and $\delta b_z$ (c), normalized to their own standard deviations, $\sigma(\delta b_i)$, and compared to their corresponding Gaussian fits (black dashed lines). Vertical black solid lines show the position of three standard deviations of each Gaussian fit. The values of $\sigma(\delta b_i)$, the limits for the Gaussian contribution, $3\sigma_G(\delta b_i)$, and the flatness, $\mathcal F_{i}$, are also indicated in each panel.}
\label{fg:pdf}
\end{center}
\end{figure*}

Figure~\ref{fg:pdf} displays the PDFs of $\delta b_x$ (a), $\delta b_y$ (b) and $\delta b_z$ (c), normalized to their own standard deviations, $\sigma(\delta b_i)$, whose values are indicated in the corresponding panels. Most of the energy is stored in the perpendicular directions, as already observed in the spectra for both perpendicular and parallel magnetic fluctuations (Figure~\ref{fg:anis_spectrum}). 
The PDFs are compared to their corresponding Gaussian fits (black dashed lines), showing the presence of fat non-Gaussian tails in each component of magnetic field fluctuations with respect to the background magnetic field, especially in the $z$ direction as expected from the higher values of $\mathcal F_{z}$ with respect to $\mathcal F_{x}$ and $\mathcal F_{y}$. The vertical black solid lines in each panel indicate the position of three standard deviations of the Gaussian fit for the corresponding magnetic fluctuations, which include 99.7\% of the Gaussian contribution. All the events that exceed these limits, $|\delta b_i|>3\sigma_G(\delta b_i)$, contribute to the non-Gaussian part of the PDFs. 
 
\begin{figure}
\begin{center}
\includegraphics [width=8.6cm]{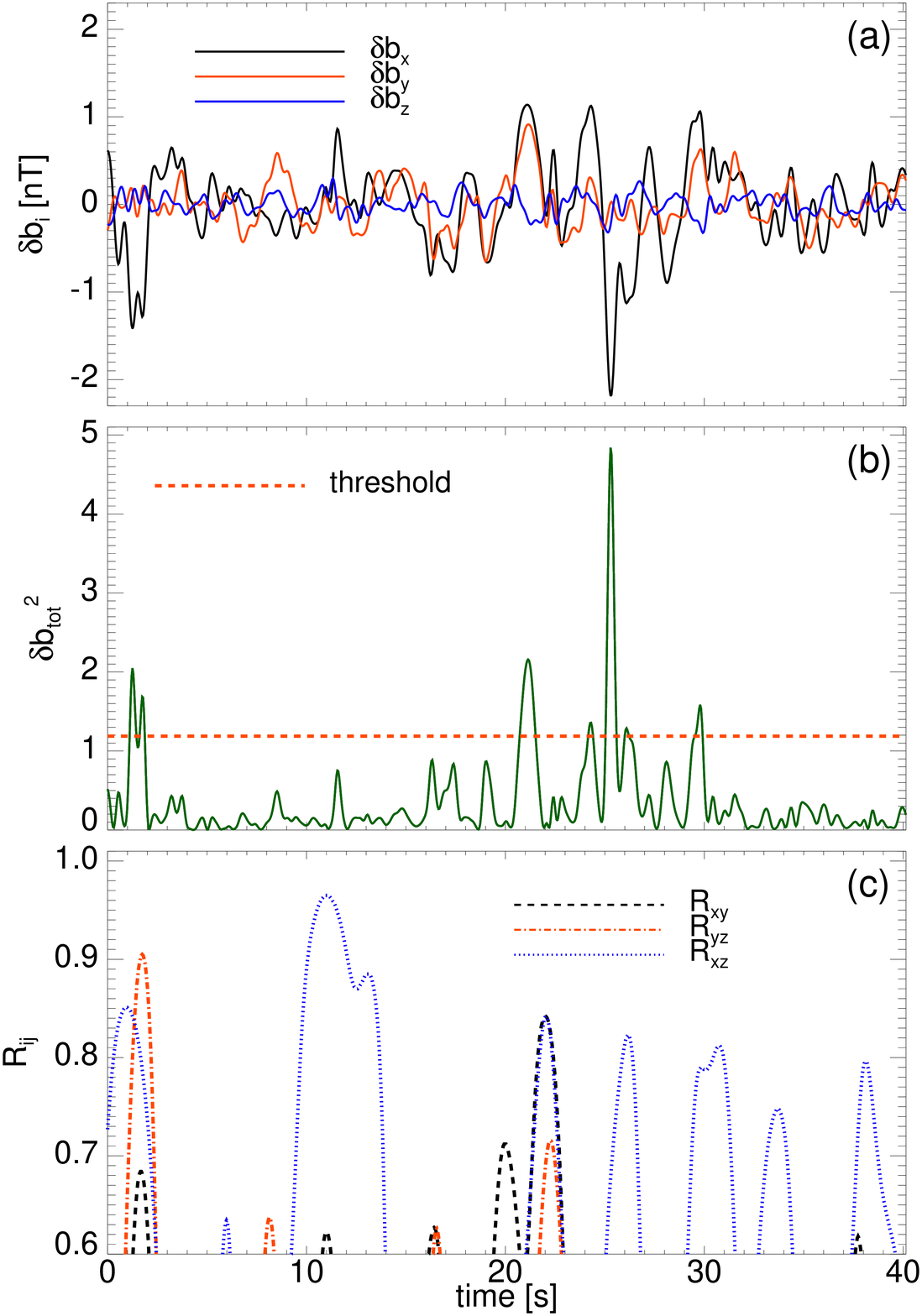}
\caption{Panel (a): Temporal evolution of turbulent magnetic fluctuations, $\delta b_i$ (see eq. \ref{eq:db})), in a considered zoom of about 40~s. Panel (b): Total magnetic energy, $\delta b_{tot}^2$, where the red horizontal dashed line indicates the threshold in the selection of intermittent events. Panel (c): Phase coherence $R_{ij}(f_0,t)>0.6$ with $f_0=2$~Hz, being the highest frequency considered for the magnetic fluctuations.}
\label{fg:confronto}
\end{center}
\end{figure}
 
To investigate the relation between the non-Gaussianity of the magnetic fluctuations and the phase coherence between the components, in Figure~\ref{fg:confronto} we consider about 40~seconds out of 40~minutes of the whole interval of fast solar wind. Panel (a) shows the time evolution of turbulent magnetic fluctuations $\delta b_x$ (black line), $\delta b_y$ (red line) and $\delta b_z$ (blue line), as defined in eq. (\ref{eq:db}). In panel (b) we display the total turbulent magnetic energy, $\delta b_{tot}^2=\delta b_x^2+\delta b_y^2+\delta b_z^2$, where the red horizontal dashed line indicates the threshold, which will be considered in the following part of the paper to select intermittent events. The threshold is defined as $3 \sigma$ of the distribution law for the amplitudes of a Gaussian vectorial field compared to $\delta b_{tot}^2$. Finally, in panel (c) we show $R_{ij}(f_0,t)>0.6$ for the highest frequency considered in the selection of intermittent events ($f_0=2$~Hz). One can see that we observe very strong peaks of coherence, i.e. magnetic field components are strongly coupled. These peaks are localized in time and very often (e.g. between 20 and 30~s) they correspond to strong peaks in magnetic energy (see panel (b)). However, sometimes strong coherence between two components, e.g. $\sim$ 10~s, corresponds to small amplitude in $\delta b_{tot}^2$, that is not selected by our threshold, that could be overestimated to select all the coherent structures.

This result verifies the link between intermittency, non-Gaussian fluctuations and phase coherence between magnetic field components, in agreement with the studies of \cite{kog07} and \cite{lio16}. Moreover, thanks to this link, we can assert that the selection of intermittent events in a turbulent signal is almost independent of the choice of a particular magnetic field component, as far as all components are coupled.

In the present paper, we select intermittent events by considering a threshold ($\sim 1.2$~nT$^2$) on the total turbulent magnetic energy, $\delta b_{tot}^2$, in the range scale $f \in [0.1,2]$~Hz, as discussed above. It is worth pointing out that the same results, that will be described in the following part of the paper, can be found if we select intermittent events by considering compressive fluctuations, $\delta b_{\parallel}=\delta |b|$, as in \cite{per16}. For the whole time interval (40 minutes) of fast solar wind, we get about 140 peaks, meaning that coherent structures cover $\sim 30\%$ of the analyzed stream (where the coherent time is evaluated as $2.5$ times the time scale of each structure, without overlapping{\footnote{Looking at each structure recovered by the selection method, we observe that coherent fluctuations are somehow larger than the time scale, defined as the time range between two minima containing a maximum of energy over the threshold. The same definition of coherent time has been used in \cite{per16}.}}). In fast solar wind the coherent structures appear to be somewhat less frequent with respect to the interval of slow solar wind described in \cite{per16}, in which $\sim 40 \%$ of the interval was covered by coherent events. A detailed analysis of the structures in fast solar wind will be presented in the following Sections of the paper.

\section{Coherent Structures}
\label{sec:structures}

To study the nature of the intermittent events, we perform a minimum variance analysis around each selected peak, identifying magnetic fluctuations which are well-localized in space and with regular profiles. These characteristics are inherent properties of coherent structures. However, since the automatic method for the selection of intermittent events recovers the most energetic peaks, it is possible that if there are few of them very close they refer to the same event. In order to avoid an overestimation of the detected events, we check all the selected peaks and we confidently identify 101 events. In particular, we find 19 isolated vortices, 32 vortex chains and 18 current sheets. For the latter, only 6 current sheets are isolated, while the other 12 are recovered at the center of vortices or sometimes at their boundaries. Moreover, for the remaining 32 structures the nature is not clear. However, no strongly compressive structures, such as solitons, magnetic holes or shocks, have been detected, in stark contrast to what is observed in slow solar wind. 

In the following, we present three examples of observed coherent structures of different nature. In Figures~\ref{fg:current},\ref{fg:alf_vor} and \ref{fg:vor_chain}, panels (a) display the modulus of the raw magnetic field measurements, namely large scale magnetic field, as observed by the four satellites (different line styles), where the FGM noise at $f>2.5$~Hz is taken off. The red double arrow indicates $\Delta \tau$, i.e. the characteristic temporal scale of the structures (see \cite{per16} for the details), while the two vertical dashed lines denote the total width of the structures ($\Delta \tau'$). 

Panels (b) show magnetic fluctuations $\delta b_i$ (with $i=x,y,z$), defined by eq.~(\ref{eq:db}), in a $BV$--reference frame which takes into account the directions of the local mean magnetic field ${\bf b_0}$ and flow velocity ${\bf v_0}$ evaluated within each structure time scale $\Delta \tau{'}$:  $z$ is aligned with ${\bf b_0}$,  ${\bf e_z} = {\bf e_b}$ (blue lines), $x$ is aligned with ${\bf v_0}$ in the plane perpendicular to ${\bf b_0}$, ${\bf e_x} = ({\bf e_b}\times{\bf e_v})\times{\bf e_b}$ (black lines), and $y$ closes the right-hand reference frame, ${\bf e_y} = {\bf e_b}\times{\bf e_x}$ (red lines). The time of each satellite is shifted taking into account the time delays with respect to C1. Moreover, panels (c) display the evolution of the current density ${\bf J}$, calculated by using the curlometer technique \cite[]{dun88,dun02}, based on four-point measurements of {\em Cluster}. The three components of ${\bf J}$ are given in the $BV$--frame, while the modulus, $|{\bf J}|$, is shown by dashed line. 

To have information on the plasma quantities, panels (d) show the behaviour of the electron density, $n_e$, evaluated by using the satellite potential \cite[]{ped95,ped01} from the Electric Field and Wave (EFW) experiment \cite[]{gus97}. These measurements have 5 vectors per second time resolution, that is much better than particle measurements on Cluster, which have 4 seconds time resolution. However, the spacecraft potential is subject to a strong spin effect as well as charging effects due to different parts of the spacecraft being illuminated as it spins. In order to have sub-spin time resolution the spin effect needs to be removed. This can be done provided the density is stable by constructing a series of phase angles for the spacecraft, and binning the corresponding potentials by angle as opposed to time. By using the median value of each bin to reduce the effects of extreme values and subtracting the mean in the interval studied the charging fluctuation can be obtained as a function of phase angle. This can be fitted with a model and subtracted from the potential measurement at each spacecraft phase angle. More details are provided in \cite{rob17}. 

Finally, panels (e) and (f) give the configuration of the four {\it Cluster} satellites in the $BV$--frame, by using different symbols and colors: black diamonds for C1, red triangles for C2, blue squares for C3 and green circles for C4. The arrows display the directions of the normal of the structures, ${\bf n}$ (black), determined by using the timing method (see Section \ref{sec:analysis}), of ${\bf v_0}$ (red) and of ${\bf b_0}$ (blue). Moreover, the black-dashed lines indicate the plane of the structures.

\subsection{Current sheet}
\label{subsec:current}

\begin{figure}
\begin{center}
\includegraphics [width=8.8cm]{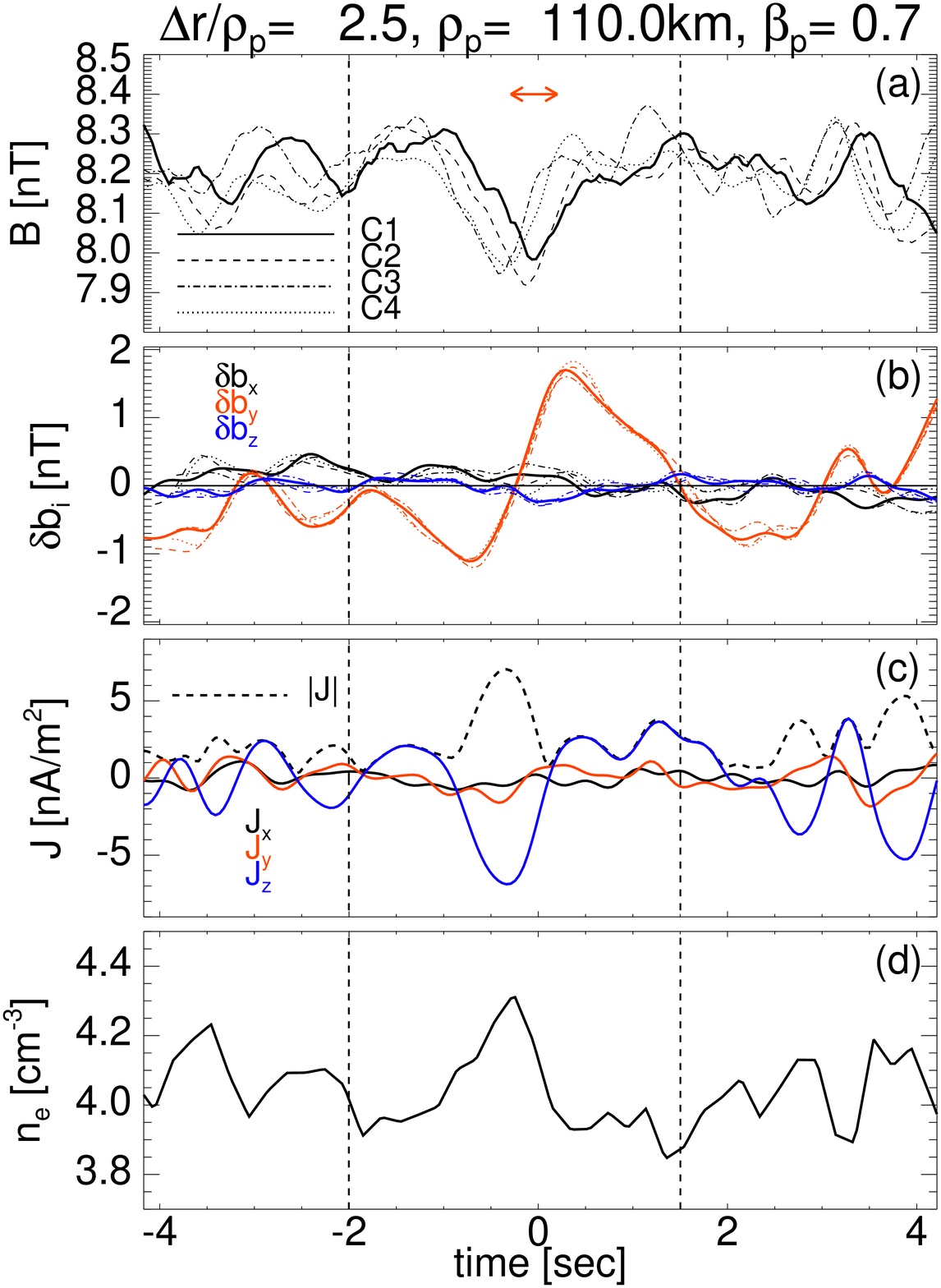}
\includegraphics [width=8.8cm]{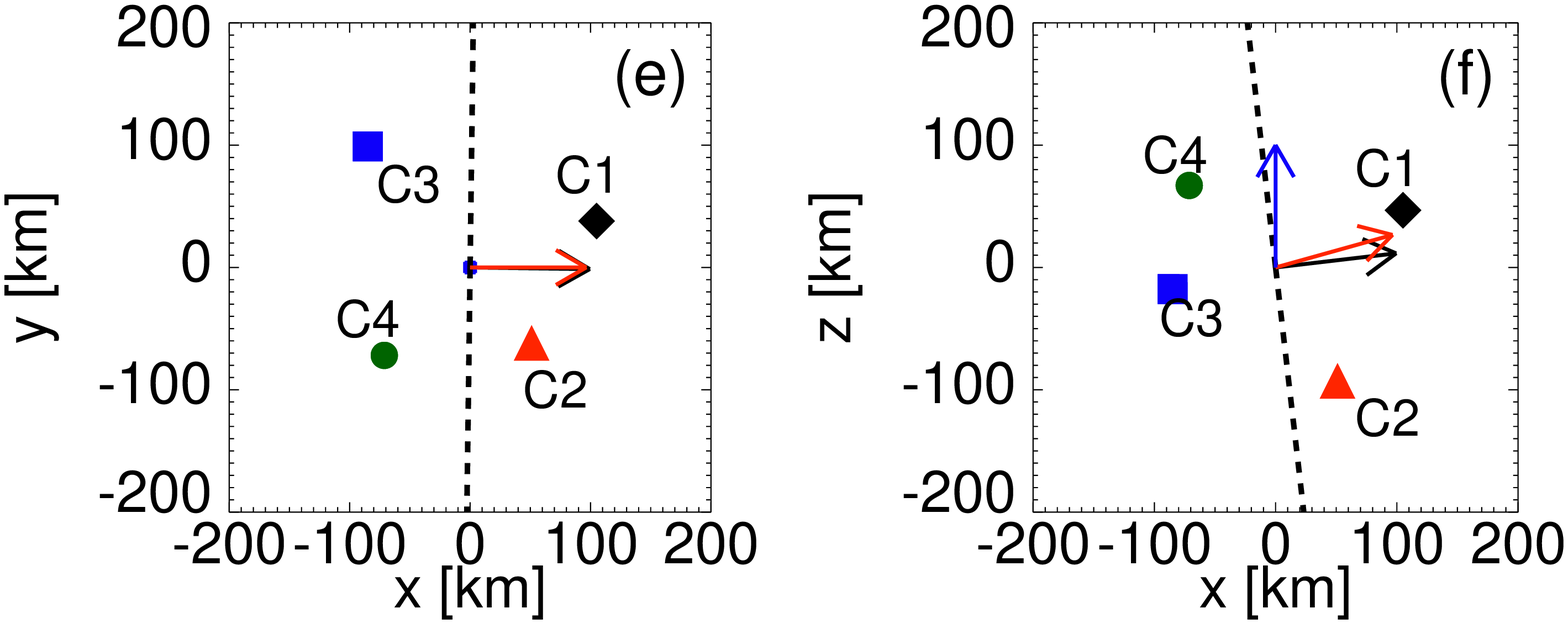} 
\caption{Example of a current sheet, centered at 18:03:36~UT. Panel (a): modulus of the large scale magnetic field observed by the four {\em Cluster} satellites (different style lines). The red double arrow indicates $\Delta \tau$, corresponding to $\Delta r$, being the characteristic scale of the structure. Panel (b): components of magnetic fluctuations defined by eq.~(\ref{eq:db}), in $BV$-frame. The time of each satellite is shifted taking into account the time delays with respect to C1. The horizontal black solid line is given as a reference for $\delta b_i=0$~nT. Panel (c): modulus (black-dashed line) and components (in $BV$-frame) of the current density. Panel(d): electron density obtained by the spacecraft potential. The vertical black-dashed lines indicate $\Delta \tau{'}$, corresponding to the total extension of the structure ($\Delta r{'}\simeq 13.5 \rho_p$). Panels (e) and (f): Configuration of  {\it Cluster} satellites in $BV$-frame: black diamonds for C1, red triangles for C2, blue squares for C3 and green circles for C4. The arrows indicate the direction of the normal (black), local flow (red) and local magnetic field (blue), while the black-dashed lines represent the plane of the structure.}
\label{fg:current}
\end{center}
\end{figure}

The first example of coherent structure is shown in Figures~\ref{fg:current}. It is an incompressible structure with a component, $\delta b_y$, which changes sign and is perpendicular to the local magnetic field. The other two components have fluctuations of very small amplitude. The reversal of the component of maximum variation is in the middle of the structure, where the large scale magnetic field has its local minimum (panel (a)) and a peak in the current is recovered (panel (c)). Minimum variance analysis applied to this structure confirms the result that is a one-dimensional (i.e. linearly polarized) alfv\'enic structure: the direction of the maximal variation ${\bf{e_{max}}}$ is perpendicular to the direction of ${\bf b_0}$ ($\theta_{max} \simeq 86^{\circ}$), while the current density is almost parallel. The four satellites observe the same amplitudes for the fluctuations, that is consistent with a planar geometry. Moreover, in the center of the structure, a peak in the density is found (panel (d)), meaning that the plasma is confined inside the structure. This event can be identified as a current sheet. Finally, panels (e) and (f) show that the normal to the structure, ${\bf n}$, is almost perpendicular to ${\bf b_0}$, while is almost parallel to ${\bf v_0}$.  
The thickness of the current sheet, estimated by using the timing method (see \cite{per16} for details), is $\Delta r \simeq 2.5\rho_p$ ($\Delta r{'} \simeq 13.5\rho_p$, being the total extension of the structure), and its velocity in the plasma frame is $\mathcal{V}_0 \simeq -(23 \pm 209)$~km/s. Therefore, it is almost convected by the flow, as expected for a current sheet.

\begin{figure}
\begin{center}
\includegraphics [width=8.2cm]{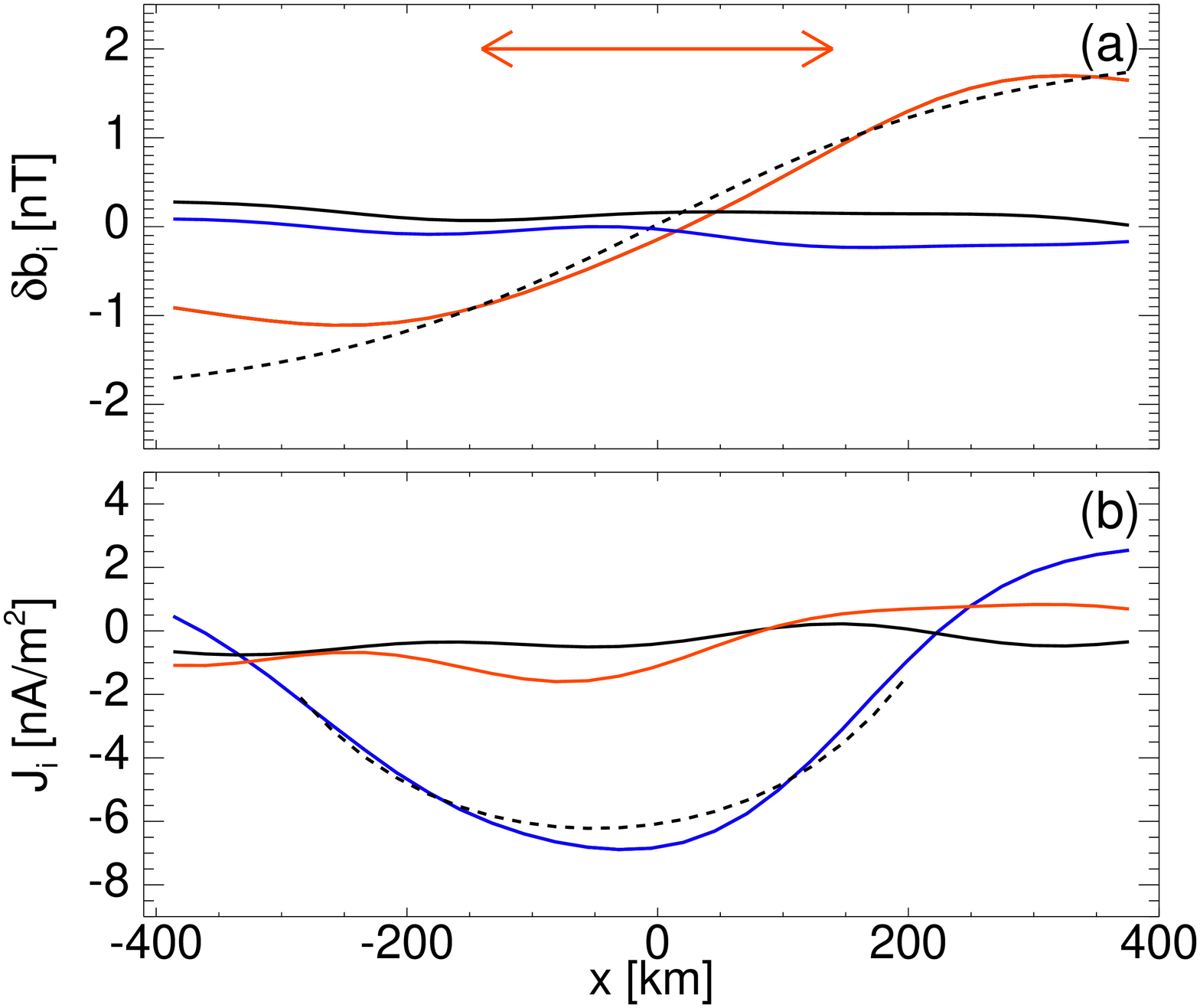}
\caption{Three components of magnetic field fluctuations (panel (a)) and current density (panel (b)) in the $BV$--frame of the discontinuity in Figure~\ref{fg:current}, as a function of the spatial coordinate, defined as $x=V*t$. The dashed lines denote the Harris profile for both magnetic field and current, while the red double arrow indicates $\Delta r$ in km.}
\label{fg:current_modello}
\end{center}
\end{figure}

A well-known one-dimensional current sheet equilibrium is the Harris current sheet \cite[]{har61}, which is a stationary solution of the Maxwell-Vlasov system. This simple model could represent analytically thin current layers at kinetic scales \cite[]{gre16}. The magnetic field profile is given by a 1D hyperbolic-tangent profile ($B = B_0 \tanh(x/L)$), where $x$ is the spatial coordinate and $L$ is the half-width of the current sheet. The corresponding profile for the current density is  $J \propto (B_0/L) \cosh^{-2}(x/L)$. Figure~\ref{fg:current_modello} shows the three components of the magnetic field fluctuations (panel (a)) and of the current density (panel (b)) in the $BV$--frame for the discontinuity in Figure~\ref{fg:current}, as a function of the spatial coordinate, defined as $x=V*t$, where $V$ is the velocity of the current sheet in the satellite frame and $t$ is the time as indicated in Figure~\ref{fg:current}. The dashed lines denote the Harris profile for both magnetic field and current, while the red double arrow indicates the characteristic scale of the structure in km, $\Delta r$. The agreement between the plasma equilibrium theory and the observed current sheet is satisfactory. The parameters of the fit for $\delta b_y$ give $B_0 \sim 2 nT$, which corresponds approximately to the level of saturation before and after the reversal of the magnetic field. Moreover, from the fit we obtain $L \sim 272$~km $=2.47 \rho_p$, while the characteristic scale of the current sheet from the timing analysis is $\Delta r \simeq 279$~km $=2.54 \rho_p$. 

It is worth pointing out that the assumptions behind the analytical model of the Harris current sheet establish that the velocity distributions for both ions and electrons must be Maxwellian. However, the considered stream of fast solar wind is characterised by anisotropic proton temperatures. Unfortunately, due to the low-resolution of ion measurements on {\it Cluster} (4 seconds), it is not possible to test anisotropic kinetic models.

\subsection{Vortex}
\label{subsec:vortex}

\begin{figure}
\begin{center}
\includegraphics [width=8.8cm]{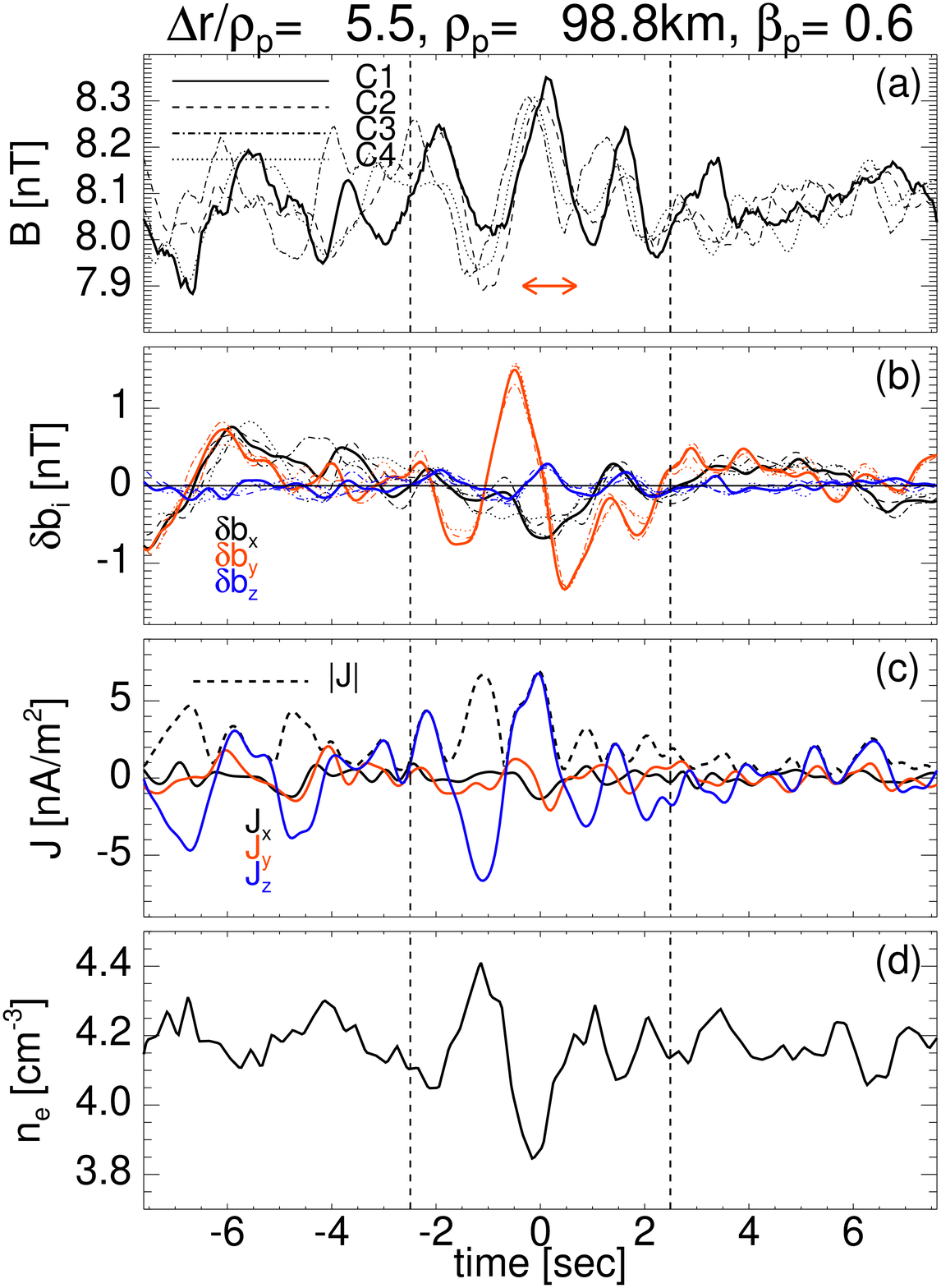}
\includegraphics [width=8.8cm]{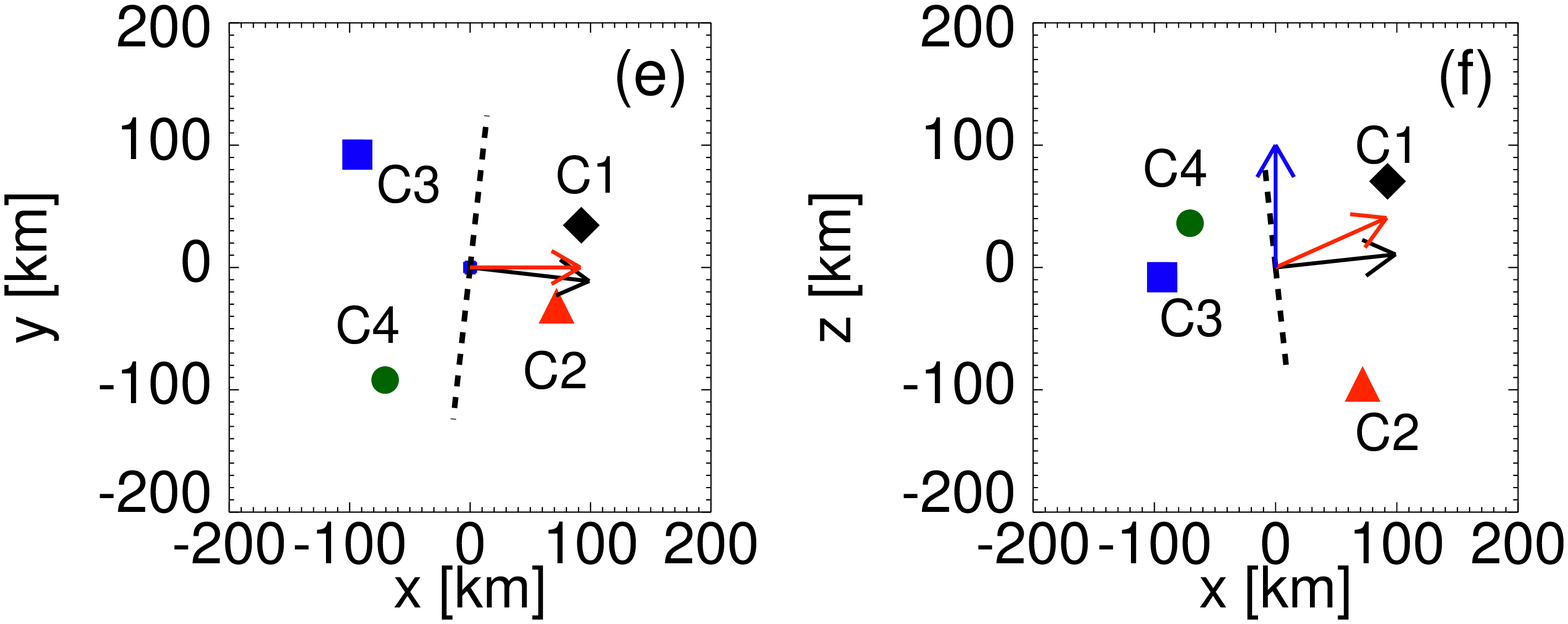}
\caption{Example of a vortex-like structure, centered at 18:06:03.6~UT and with $\Delta r{'} \simeq 27 \rho_p$. The panels are the same as in Figure~\ref{fg:current}.}
\label{fg:alf_vor}
\end{center}
\end{figure}

Another example of coherent structure observed in this stream of fast solar wind is shown in Figure~\ref{fg:alf_vor}. The background magnetic field (panel (a)) is characterized by a modulated fluctuation, observed by the four spacecraft, with a local maximum in the middle of the structure. The corresponding fluctuations, $\delta b_i$, are more localized, with the principal variation in the plane perpendicular to the local magnetic field, $\delta b_y$, and small compressive fluctuations, $\delta b_z \ll \delta b_y$. Moreover, a minimum variance analysis indicates that the intermediate component is not negligible, i.e. the event is a bi-dimensional structure, and both the directions of maximum and intermediate variance are in the plane perpendicular to ${\bf b_0}$ ($\theta_{max} \simeq 88^{\circ}$ and $\theta_{int} \simeq 87^{\circ}$), while the direction of minimum variance is along ${\bf b_0}$ ($\theta_{min} \simeq 3^{\circ}$). 
Panel (c) displays the current density, ${\bf J}$, that is mainly in the direction parallel to ${\bf b_0}$, while panel (d) shows the electrons density, that exhibits a fluctuating behaviour ($\delta n_e \sim 0.1$~cm$^{-3}$) and it is anti-correlated with the background magnetic field, with a local minimum in the center of the structure. Finally, panels (e) and (f) show the result of the timing method for the normal of this structure, that is almost perpendicular to ${\bf b_0}$ ($\theta_{nB} \simeq 84^{\circ}$), while is almost parallel to ${\bf v_0}$. The velocity of propagation along the normal and in the plasma frame is $\mathcal{V}_0 \simeq -(27\pm240)$~km/s. The characteristic scale for this two-dimensional structure is $\sim 5.5\rho_p$, while the total width, corresponding to $\Delta \tau'$, is $\sim 27 \rho_p$. 

The structure looks like a monopolar Alfv\'en vortex \cite[]{pet92,ale08_n}, crossed by the four satellites more or less at the same distance from the center. In general, a monopolar vortex is a tubular structure which is aligned with the magnetic field direction and is a nonlinear analytical solution of the ideal, incompressible MHD equations. However, in this case, the structure shows a pressure balance, i.e. an anti-correlation between density and magnetic field, not predicted by the incompressible model. To verify the vortex topology of this structure, we fit the observed fluctuations with the analytical model for a monopolar Alfv\'en vortex, derived from the vector potential, ${\bf A}$ \cite[]{pet92,ale08_n,rob16}. The longitudinal current is given by ${\bf J}=\nabla_{\perp}^2 {\bf A}$.

\begin{figure*}
\begin{center}
\includegraphics [width=9.0cm,angle=90]{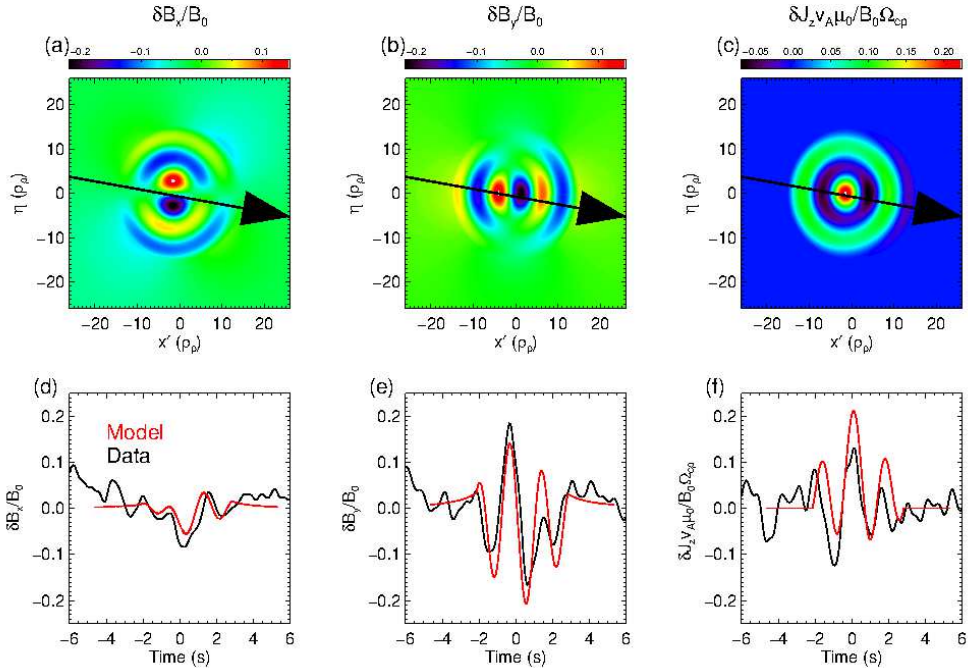}
\caption{Quasi-monopolar Alfv\'en vortex model for the fluctuations in Figure~\ref{fg:alf_vor}. Panels (a) and (b) show the perpendicular fluctuations of the vortex solution $\delta b_{x}(x',\eta)$ and $\delta b_{y}(x',\eta)$, respectively, normalized to the local mean magnetic field, ${\bf b_0}$, while panel (c) shows the normalized longitudinal current, $\delta J_{z}(x',\eta)$. The spatial dimensions are given in units of proton Larmor radius, $\rho_p$. The arrows denote the paths of the spacecraft through the vortex which give the modeled fluctuations (red lines) in panels (d)--(f).}
\label{fg:vortex_model}
\end{center}
\end{figure*}

The vortex is modeled in the $x'-\eta$ plane which are two directions perpendicular to the vortex axis and the $x'$ axis makes an angle of {\bf $10^{\circ}$} with the relative path of the spacecraft through the plasma. Moreover, the angle of the vortex axis with respect to the magnetic field direction is $2^{\circ}$. The vortex is modeled with a constant amplitude of $A_{0}=-0.3$ and the vortex diameter is set at $30\rho_{p}$. This is motivated by the value obtained from timing analysis and is much larger than the inter-spacecraft distances, consistent with all spacecraft seeing similar fluctuations. Finally, the impact parameter, i.e. the distance from the center at $x'=0$ is $-0.05a$, being $a$ the radius of the vortex.

Panels (a) and (b) of Figure~\ref{fg:vortex_model} show the perpendicular fluctuations of the vortex solution $\delta b_{x}(x',\eta)$ and $\delta b_{y}(x',\eta)$, respectively, normalized to the local mean magnetic field, ${\bf b_0}$, while panel (c) shows the normalized longitudinal current, $\delta J_{z}(x',\eta)$. The spatial dimensions are given in units of the proton Larmor radius, $\rho_p$. The relative path of the virtual spacecraft is denoted by the arrows. The analytical solution of the Alfv\'en vortex and {\it Cluster} data are compared in panels (d)--(f) of Figure~\ref{fg:vortex_model} and show a good agreement for both magnetic components and current density.

We observe 19 structures similar to the example in Figure~\ref{fg:alf_vor}. Moreover, only 2 of them present a compressive nature, where the ratio between the parallel and perpendicular magnetic field fluctuations is higher than 0.35 \citep{per16}. 

In order to figure out on the possible nature of these structures, we perform an analysis on the polarization \citep{he11,he12,tel15}, though here we are studying structures and not waves (we have not a specific frequency for them, but a range of scales is covered). For all the vortices (except for one) we find, in the satellite frame and in the ($\delta b_y$--$\delta b_z$) plane perpendicular to the normal, an elliptical polarization with the major axis perpendicular to the local mean magnetic field, right-handedness with respect to the direction of the normal (that in the wave approximation represents the ${\bf k}$ direction). This result is consistent with previous studies of the dissipation range \cite[e.g.]{gol94} and with the fact that in the case of an Alfv\'en wave, the increase of the angle between ${\bf k}$ and ${\bf B}_0$ produces a change in the polarization from left- to right-handed polarization \citep{gar86}. 

However, by considering the polarization in the plasma frame (i.e. by considering the sign of $\mathcal{V}_0$), the result changes. If the velocity of the structure is smaller than the velocity of the solar wind along the normal, the propagation is antiparallel to the normal, so the observed polarization is inversed in the plasma frame (i.e. left-handed polarization). Since the observed vortices have both positive and negative $\mathcal{V}_0$, the polarization can be both right- and left-handedness in the plasma frame. Anyhow, sometimes the value of $\mathcal{V}_0$ can be very small and/or its error be important. Therefore, a definitive conclusion is very difficult.

Finally, we compare the ratio between the parallel and the total magnetic energy with what expected for Kinetic Alfv\'en Waves \citep{bol13}. In the case of the vortices the ratio is very low, i.e. about one order of magnitude lower than expected for Kinetic Alfv\'en Waves.

\subsection{Vortex chain?}
\label{subsec:chaos}

\begin{figure}
\begin{center}
\includegraphics [width=8.8cm]{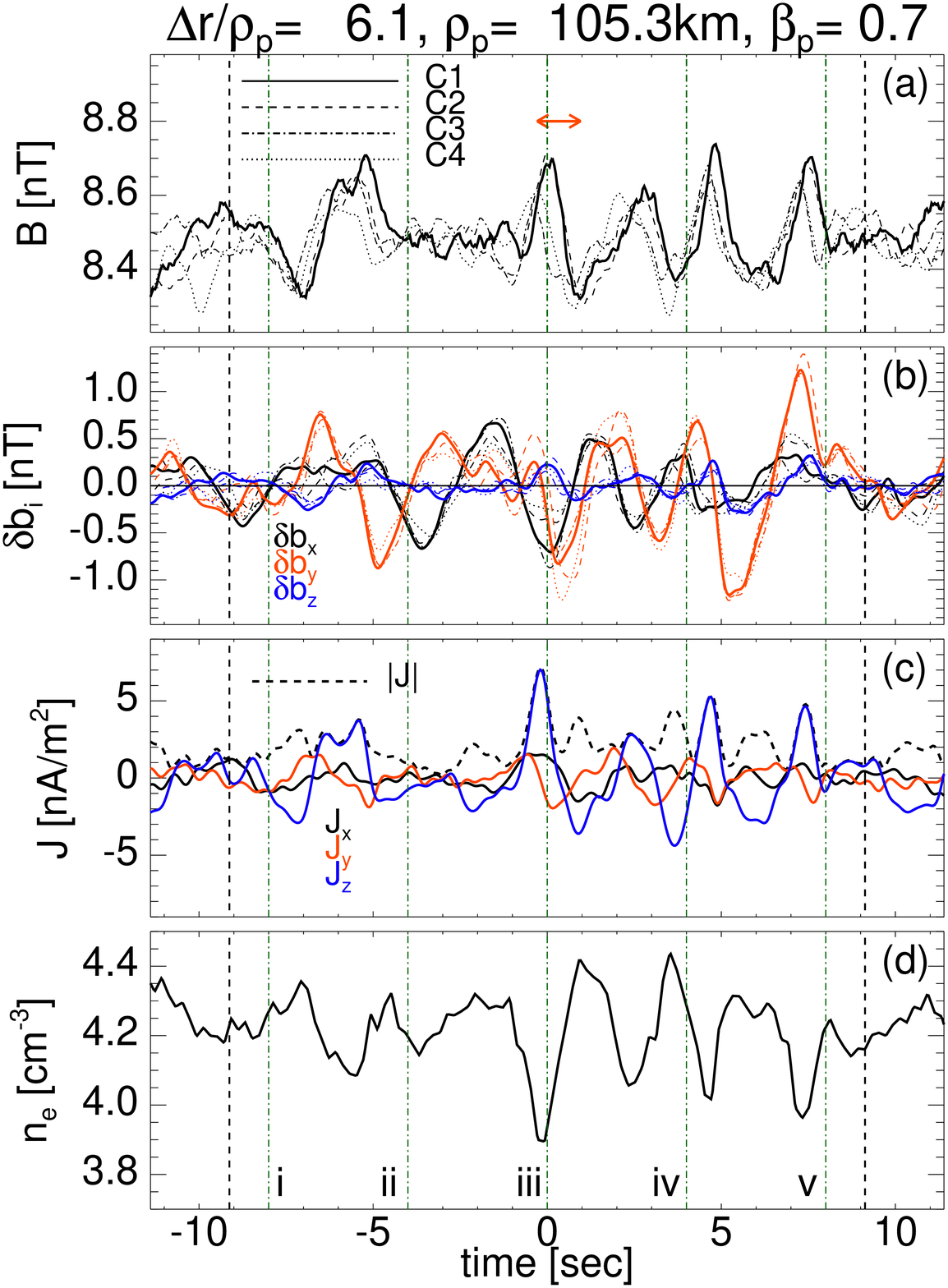}
\includegraphics [width=8.8cm]{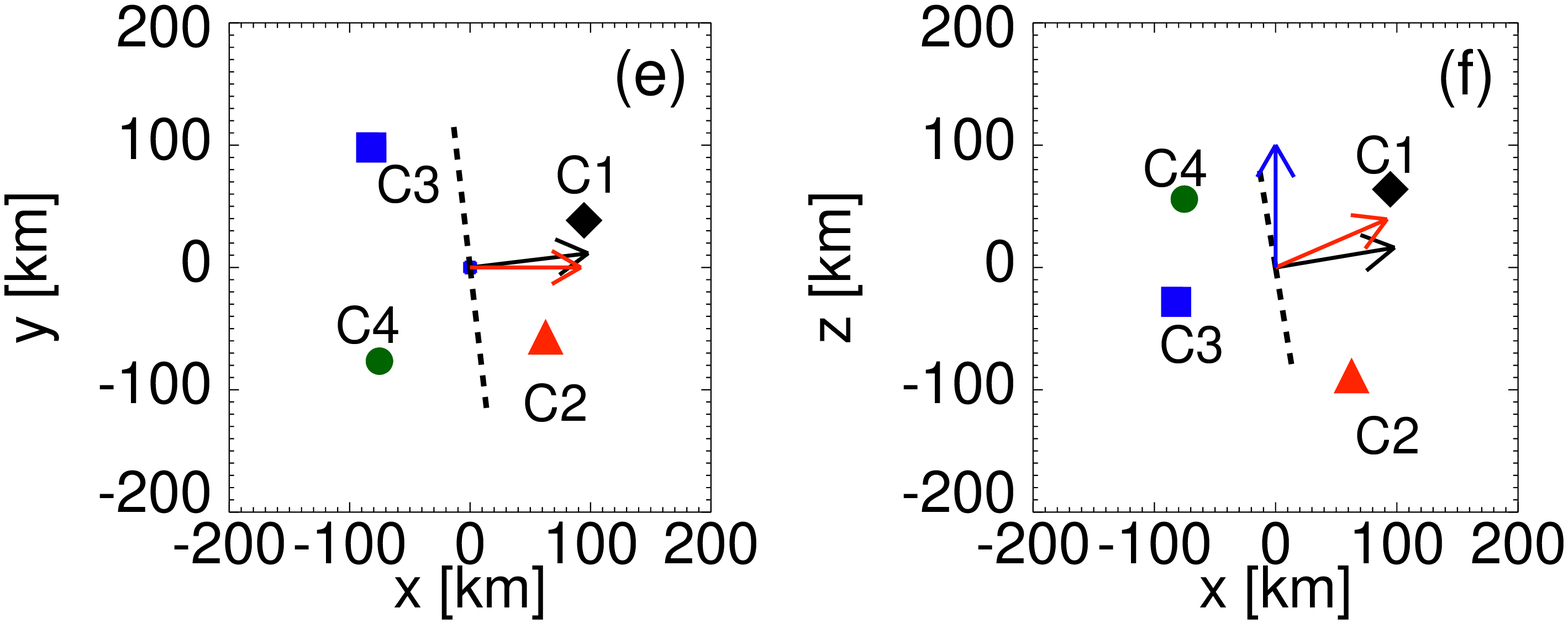}
\caption{Example of an interaction of Alfv\'en vortices, centered at 17:37:33.6~UT and with $\Delta r{'} \simeq 89 \rho_p$. The panels are the same as in Figure~\ref{fg:current}. Moreover, the vertical green dot-dashed lines in each panel denote the time of the electron pitch angle distributions in Figure~\ref{fig:blah}.}
\label{fg:vor_chain}
\end{center}
\end{figure}

The last example of observed structure is given in Figure~\ref{fg:vor_chain}. The large scale magnetic field is characterized by different oscillations and the same behaviour is observed in the components of the magnetic field fluctuations in panel (b), showing significative oscillations in the plane perpendicular to ${\bf b_0}$ ($\theta_{max} \simeq 88^{\circ}$ and $\theta_{int} \simeq 84^{\circ}$), while the minimum variance direction is almost parallel to ${\bf b_0}$ ($\theta_{min} \simeq 7^{\circ}$) as the current density. Moreover, the principal spatial gradient is $\nabla _{\perp} \gg \nabla_{\|}$, which gives ${\bf n} \perp {\bf b_0}$ (see panels (e) and (f)). Furthermore, the normal of the structure is almost parallel to ${\bf v_0}$ and the velocity of propagation in the plasma rest frame is $\mathcal{V}_0 \simeq -(52\pm580)$~km/s. The characteristic scale for this structure, indicated by the red double arrow, is $\sim 6.1\rho_p$, while the total width is $\sim 89 \rho_p$. Finally, as in the case of the vortex in Figure~\ref{fg:alf_vor}, the electron density (panel (d)) is anti-correlated to the large scale magnetic field, meaning that also this event is in pressure balance.

\begin{figure*}
\begin{center}
\includegraphics [width=17.5cm]{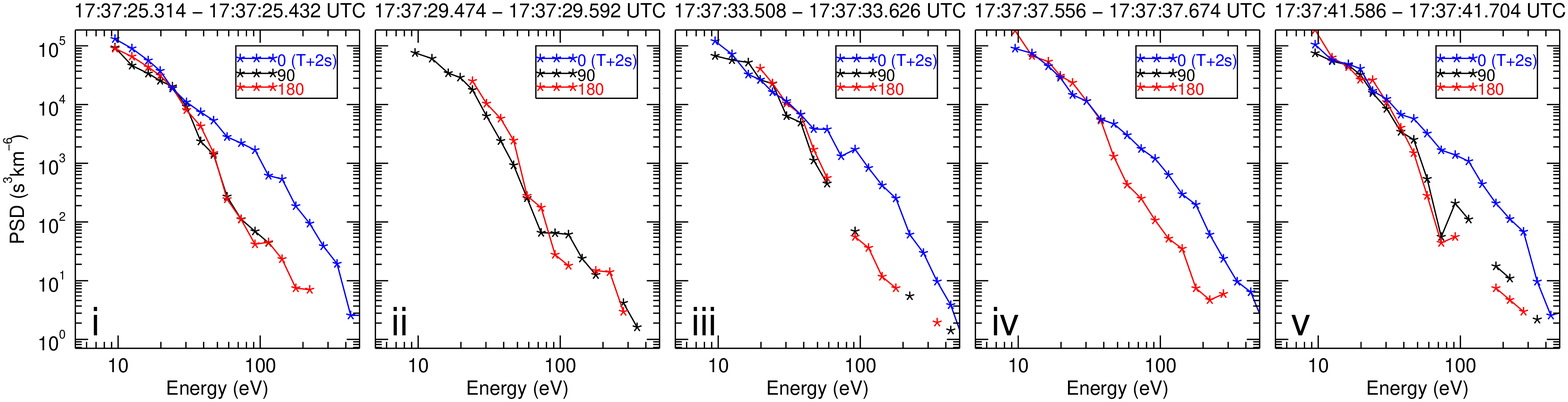}
\caption{Five incomplete pitch angle distributions, taken during about 20 minutes surrounding the centre of the vortex chain in Figure~\ref{fg:vor_chain}. Each panel shows a cut of the electron distribution at pitch angles $0^{\circ}$ (blue line), $90^{\circ}$ (black line) and $180^{\circ}$ (red line). The time of the cuts corresponds to the vertical green dot-dashed lines in Figure~\ref{fg:vor_chain} for pitch angles $90^{\circ}$ and $180^{\circ}$, while pitch angle $0^{\circ}$ is taken 2~s later. No coverage of pitch angle $0^{\circ}$ is available in panel ii, while in panel iv the same problem is found for pitch angle $90^{\circ}$}
\label{fig:blah}
\end{center}
\end{figure*}

Figure~{\ref{fig:blah}} shows five incomplete electron pitch angle distributions, taken during about 20 seconds surrounding the center of the structure. In particular, these plots are cuts of the distribution at pitch angles  $0^{\circ}$ (blue line), $90^{\circ}$ (black line) and $180^{\circ}$ (red line). Some gaps in the lines are due to missing data. Moreover, the electron pitch angle distributions are not corrected from the spacecraft potential also because of missing data. However, the spacecraft potential is less than 9~eV since this is the lowest energy PEACE measures and no photoelectrons are visible.

In its most common operating mode, the PEACE electron spectrometer \cite[]{joh97, faz09} returns a 2D pitch angle distribution from one or both of its two sensors every spacecraft spin (i.e. $\sim$4\,s). Each pitch angle distribution is constructed from two energy sweeps taken two seconds apart, when a sensor's field of view is looking along and against the magnetic field direction, respectively. An individual sweep is typically completed in $\sim$0.125\,s. Thus, it is possible to examine the electron properties of small-scale structures using PEACE, albeit without complete pitch angle coverage, by considering an individual sweep that was taken during the passage of that structure over the spacecraft. 

In Figure~{\ref{fig:blah}} the time of the cuts corresponds to the vertical green dot-dashed lines in Figure~\ref{fg:vor_chain} for pitch angles $90^{\circ}$ and $180^{\circ}$, while pitch angle $0^{\circ}$ is taken 2~s later.
We observed a typical strahl signature for pitch angle $0^{\circ}$ in each panel of Figure~{\ref{fig:blah}}, meaning that it does not change over all the considered 20 minutes. No information is available in panel ii due to the loss of coverage.

Let's consider now the electron distributions for pitch angles $90^{\circ}$ and $180^{\circ}$. Panels iii and iv of Figure~\ref{fig:blah} are from two sweeps taken at the center of the structure ($t=0$) and 4 seconds later ($t=+4$~s), respectively. At these times, that correspond to the vortex central region, the electron distributions seem to be typical solar wind distribution: almost isotropic with a spectral break between the core and the halo at about $60$~eV. No evidence of accelerated particles or beams is observed. Moreover, in panel iv there is no data from pitch angle $90^{\circ}$ because at that time the magnetic field direction changes significantly, losing coverage.

A different situation is found close to the vortex boundary ($t=-4$~s and $t=+8$~s), where the theoretical current model for a vortex presents a discontinuity. Here, in panels ii and v, the electron distributions are atypical, with an increase in the phase space density, localized in energy at $\sim 100$~eV, in both antiparallel and perpendicular electrons. These distributions could be unstable and generate waves, such as Langmuir waves. Finally, panel i shows the electron distribution at $t=-8$~s, that is close to a current sheet or another vortex boundary and is slightly atypical, with an increase in the phase space density around $\sim 100$~eV for pitch angle $180^{\circ}$. 

The structure in Figure~\ref{fg:vor_chain} could be interpreted as a chain of three adjacent vortices, crossed by the spacecraft at different distance from the center of each vortex. It is interesting that, within the vortices, the electron distribution functions are typical solar-wind distributions but, close to their boundaries, electrons beams are observed at pitch angles $90^{\circ}$ and $180^{\circ}$, while in pitch angle $0^{\circ}$ we do not see any changes. However any signature of a similar amplitude to those observed in pitch angles $90^{\circ}$ and $180^{\circ}$ would be small compared to the signature of the strahl. Unfortunately, due to the low time resolution of the measurements, we are not able to follow the evolution of the electron pitch angle distribution in each point of the structure. Even worse is the case of ion measurements, whose time resolution on the considered solar wind stream is so low ($\sim 1$ minute in the low geometric factor side of the HIA instrument for solar wind measurements) that no detailed information are available and also the angular resolution is still quite low to study the deformation of the velocity distribution (B. Lavraud 2017, private communication).

On the other hand, the structure could be also described as an interaction between vortices.
By assuming, for example, that the two initial vortices are centered around $-5$ and $5$~s with an isolated extension of about 8~s each, we could expect that the center of the total structure corresponds to the region of interaction. From panel (c), the signs of the vortex currents suggest that the Lorentz force could attract the two vortices. Moreover, looking at the shape of the magnetic fluctuations in panel (b), it seems that the vortex on the left side is more distorted with respect to the right one. This could be due to different and opposite velocities of the vortices (higher for the vortex on the right side) or to the different initial amplitudes of the vortices, that produce a different level of distortion at this stage of the interaction. This stage, in fact, could be only a transition phase for this interaction between vortices and this configuration could collapse later in a single larger vortex \cite[]{nov79}.

\section{Multi-satellite analysis}
\label{sec:analysis}

The Cluster mission provides a unique opportunity to determine the three-dimensional, time-dependent characteristics of small-scale structures, using four-point measurements given by identical instrumentation on the four satellites. Indeed, multi-spacecraft observations exhibit a connection between space and time: the same physical observables are measured not only at different points in space, but also at different instants in time. To exploit this opportunity, we use the {\it timing} method \cite[]{sch98, per16} to characterize the coherent structures observed in this interval of fast solar wind. 

The timing method is based on time and space separations and allows to determine the velocity, $\mathcal{V}$, and the direction of propagation, ${\bf n}$, of a locally planar structure moving with a constant speed in the spacecraft frame. All the details about the method and the conditions of its validity can be found in Section 4.2 of \cite{per16}. In the present work, the timing technique allows us to study only a subset of 33 structures (out of 101), for which the method keeps validity and we are sure to properly determine ${\bf n}$ and $\mathcal{V}$. 

\begin{figure}
\begin{center}
\includegraphics [width=6.8cm]{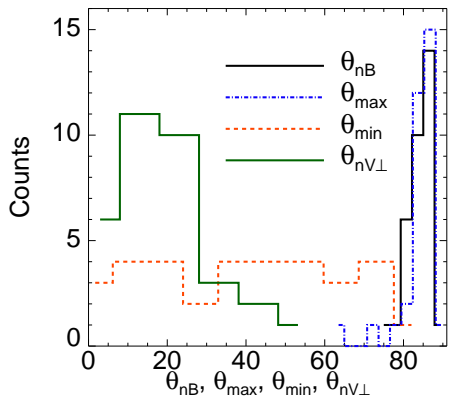}
\caption{Distributions of the angles of the local magnetic field ${\bf b_0}$ with (i) the normal ${\bf n}$ ($\theta_{nB}$, black solid line); (ii) the direction of maximal variance ${\bf e_{max}}$ ($\theta_{max}$, blue dot-dashed line) and (iii) minimum variance ${\bf e_{min}}$ ($\theta_{min}$, red dashed line); and of the angle between ${\bf n}$ and the local solar wind velocity in the (x, y)-plane of the $BV$--reference frame ($\theta_{nV_{\perp}}$, green solid line).}
\label{fg:direction}
\end{center}
\end{figure}

Figure~\ref{fg:direction} shows the distributions of the angles between the local magnetic field ${\bf b_0}$ and several vectors, i.e. the normal of the structure, the directions of maximal and minimum variance, and the distribution of the angle between ${\bf n}$ and the local solar wind velocity in the perpendicular plane, to statistically characterize the geometry and the properties of the observed coherent structures. In particular, $\theta_{nB}$ (black solid line) is always close to $\sim 90^{\circ}$, meaning that all the observed coherent structures have a perpendicular wave-vector anisotropy ($k_{\perp} \gg k_{\parallel}$), while $\theta_{nV_{\perp}}$ (green solid line) is peaked around $\sim 20^{\circ}$, where $V_{\perp}$ is the local solar wind velocity in the (x, y)-plane of the $BV$--reference frame. These results are in agreement with the case of slow solar wind studied by \cite{per16}. On the other hand, ${\bf e_{min}}$ (red dashed line), the angle between ${\bf b_0}$ and the direction of minimal variance, shows an uniform distribution in the range between $0^{\circ}$ and $80^{\circ}$. Moreover, $\theta_{max}$ (black solid line), the angle between ${\bf b_0}$ and the direction of maximal variance, is almost peaked around $\sim 90^{\circ}$, emphasizing the absence of compressive structures in fast solar wind. Here, the structures are mostly Alfv\'enic ($\delta b_{\perp} \gg \delta b_{\|}$) with very small compressive component with respect to slow solar wind structures. These results, in comparison with the analysis done by \cite{per16}, suggest a different nature, in terms of features of coherent structures, between fast and slow streams of solar wind. 

\begin{figure*}
\begin{center}
\includegraphics [width=8.8cm]{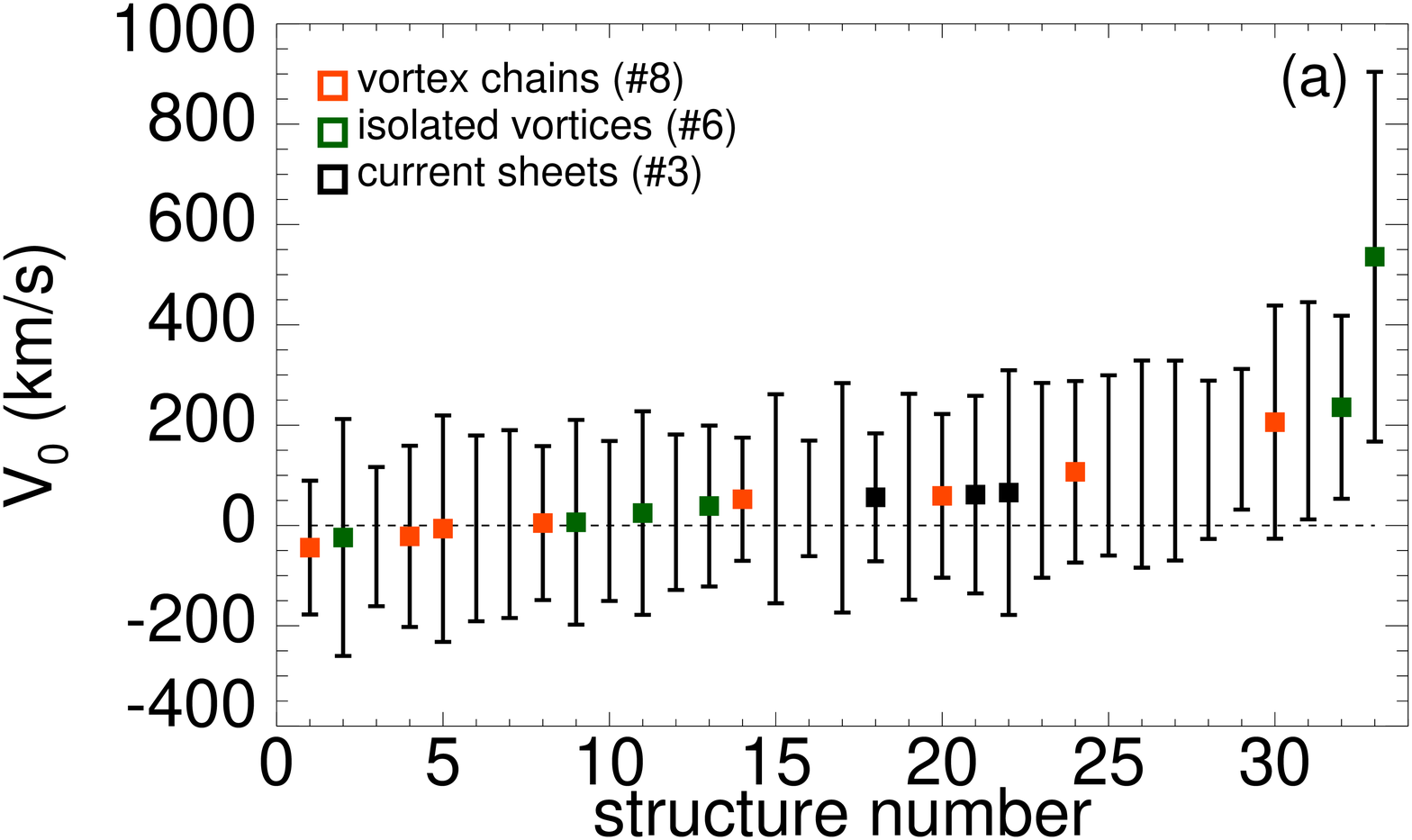}
\includegraphics [width=6.cm]{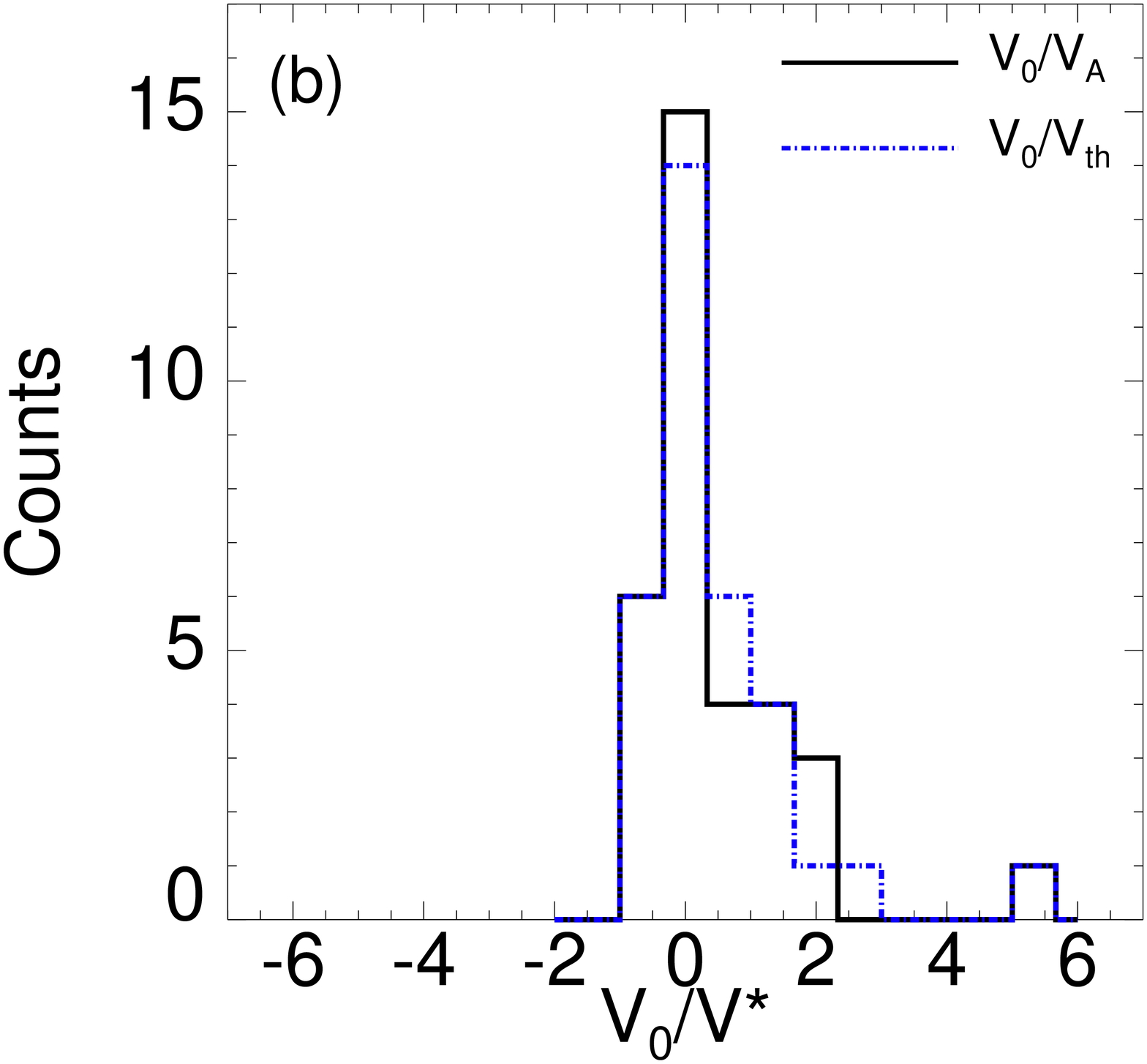}
\caption{Panel (a): Velocity of 33 structures in the plasma rest frame, $\mathcal{V}_0$, with the corresponding error bars, arranged in increasing order of the $\mathcal{V}_0$ values. The different kinds of the structures are indicated by different colours and the details can be find in the legend. Panel (b): Distribution of $\mathcal{V}_0$ normalized to the Alfv\'en speed, $V_A$ (black solid line) and to the proton thermal speed, $V_{th}$ (blue dot-dashed line).}
\label{fg:speed}
\end{center}
\end{figure*}

To study more in detail the differences between the two streams, in Figure~\ref{fg:speed}, we statistically investigate the velocity of the structures along the normal direction in the plasma frame, $\mathcal{V}_0 = \mathcal{V} - {\bf v}_{sw} \cdot {\bf n}$, being ${\bf v}_{sw}$ the local mean speed of the solar wind, and the corresponding error, $d\mathcal{V}_0$. For the details, please, refer to Section 4.2.2 of \cite{per16}. It is worth pointing out that $d\mathcal{V}_0$ is larger in the case of fast solar wind than in the slow wind case because is proportional to the value of the wind speed (the error on the solar wind velocity is estimated as the 5\% of the value of the speed). Figure~\ref{fg:speed}(a) shows $\mathcal{V}_0$ for the 33 structures, ordered by its increasing value, with the corresponding error bars. Contrary to what is found for the case of slow solar wind, in which the 25\% of the observed structures had significant velocities different to zero \cite[]{per16}, in the present stream of fast solar wind all the structures are simply convected by the flow. Only a clear example of no convected structure is observed, but it is probably due to the incertitude in the identification of  the range of localization. Moreover, also the distribution of $\mathcal{V}_0$, normalised to the Alfv\'en speed, $V_A$ (black solid line) and to the proton thermal speed, $V_{th}$ (blue dot-dashed line) in panel (b) of Figure~\ref{fg:speed} suggests the same result, where the characteristic velocities are calculated in the upstream region for each structure, known from the sign of $\mathcal{V}_0$. Both the distributions are peaked around $0$ and vary between $[-1,2] V_A$ or $V_{th}$. No fast structures are found, contrary to the case of slow solar wind \cite[]{per16}. Only an event of very fast structure ($\mathcal{V}_0 \sim 6 V_A$) is observed, corresponding to the no convected structure in panel (a), that in any case does not have a statistical meaning. 

\section{Conclusions and Discussion}  
\label{sec:conclusion} 

In this paper, we have investigated the nature of magnetic turbulent fluctuations around ion scales in a stream of fast solar wind, by using high-resolution \textit{Cluster} data. The results are complementary to the recent statistical study in slow wind plasma by \cite{per16}.  

The study of the distribution of energy in time and frequencies shows the presence of localized regions in time that cover a certain range of scale. The same non-homogeneous distribution of energy localized in time, covering certain frequencies, was already highlighted in slow solar wind \cite[]{per16}. Thus, independently of the streams, the solar wind ion scales appear to be characterized by a strong intermittency which could play an important role in dissipation and particle energization. 

A detailed study of magnetic fluctuation in the range $f \in[0.1,2]$~Hz has shown that this region is also characterized by a high phase coherence between the magnetic components and by a significant non-Gaussianity. The departure from Gaussianity of the turbulent fluctuations reflects a non-homogeneous (intermittent) distribution of the turbulent energy, with the appearance of structures characterized by a finite degree of phase synchronization. Therefore, intermittency, phase coherence and non-Gaussian fluctuations are found to be strictly related, in agreement with previous studies \cite[]{kog07,lio16}. 

We show that, at ion scales, the observed intermittency is related to convected coherent structures with a strong wave-vector anisotropy in the perpendicular direction with respect to the local magnetic field ($k_{\perp} \gg k_{\parallel}$). In particular, the fast solar wind appears to be dominated by Alfv\'en vortices (isolated or in chains), with small compressive part (for most of them the ratio between the parallel and the total magnetic energy is less than 10\%), and by several current sheets aligned with the local magnetic field, convected by the flow. These results are in agreement with a recent analysis by \cite{lio16} in a stream of fast wind observed by {\it Wind} spacecraft. The authors found the presence of Alfv\'en vortex-like structures and current sheets, that drastically contributes to the spectral shape of the magnetic field spectrum at ion scales. Furthermore, the comparison of this stream of fast solar wind with the results described in \cite{per16} in slow wind context suggests that the latter one is much more complex, with the presence also of strongly compressive structures, such as magnetic holes, solitons and shocks, with smaller amplitudes with respect to the Alfv\'enic structures, but that propagate in the plasma rest frame.  

In a separate study using the MSR technique, \cite{rob17} studied intervals of fast and slow wind plasma. In the fast wind phase speeds were found to be very slow often below the Alfv\'en speed for incompressible and compressible magnetic fluctuations. Meanwhile in the slow wind the compressible magnetic field showed evidence of fast propagating fluctuations, with some faster than the magnetosonic speed. This is consistent with the results discussed in both \cite{per16} and in the present paper. Therefore, slow solar wind presents a more complex physics with respect to fast wind, where fast structures, moving in the flow, could lead to the generation of some instabilities with additional effects on particles \cite[]{pap72}. 

The difference in the observed families of structures in slow and fast streams fits also into a more general context of the source of these winds \cite[]{fel05}. For example, the presence of compressive structures in slow solar wind could be the result of the interaction of the wind with the heliospheric current sheet \cite[]{bur02,bro15} or be remnant features of large scale coronal structures \cite[]{wil67,mcc00}. Thanks to Solar Orbiter and Parker Solar Probe, it will be possible to measure the turbulent and structured electric and magnetic fields associated with shocks, reconnection and stochastic energization in unexplored plasma environments, investigating plasma physics in the source regions of both fast and slow wind. Moreover, Solar Orbiter and Parker Solar Probe will give access to the heliocentric variation of the turbulence, allowing the study electromagnetic field fluctuations and particle energization processes as a function of radial distance. 

Understanding intermittency phenomenon and the related formation of small-scale coherent nonlinear structures could provide key insights into the general problem of dissipation in collisionless plasma and more particularly in solar wind. The physics of dissipation strongly depends on the different family of structures because of the different physical processes involved in the generation and/or evolution of the considered coherent structure. Our results (in the present paper for fast solar wind and in \cite{per16} for slow solar wind) show that the vortex-like structures are the dominant form in the observed structures; thus, they may play a major role in the dynamics of the solar wind plasma at ion scales. A very recent theoretical study in nearly incompressible magnetohydrodynamic (NI MHD) turbulence by \cite{zan17} has shown that two-dimensional vortex structures are explicitly predicted by the model. In particular the NI MHD formulation describes the transport of majority 2D, and minority slab, turbulence throughout the solar wind. This result supports the fact that in both fast and slow solar wind the vortex-like structures are the most frequent intermittent events. It is worth pointing out that the model for NI MHD turbulence deals with the fluid range while the present paper deals with coherent structures at kinetic scales. However, as far as coherent structures cover a very large range of scales, the detected strong events at ion scales show a waveforms covering $\sim 30 \rho_p$ (the diameter of the vortices). Therefore, the NI MHD turbulence model could be still applicable.

The observed vortices can be divided in two subfamilies with different properties. In particular, we found strongly localized vortices as well as vortex chains. In the first case, localized vortices could trap particles and, propagating in the flow, could excite density fluctuations and increase heat and mass transport processes. This is the scenario of a strong vortex turbulence \cite[]{abu09}. However, these isolated vortices could be the result of merging processes from smaller to larger scale structures, like as in the mechanism for self-organization in ideal fluids, creating a finite number of large, well separated vortices \cite[]{nov79,mcw84,bra00}. This phenomenon is called {\em vortex collapse} and one could expect similar phenomena to occur also in plasmas. For localized structures, as long as the mutual distance between the vortices is larger than their size, there is not interaction between them and they can be described as Alfv\'en-type vortices \cite[]{pet92}. However, when the vortices are closer together their shapes start to deform. The vortex merger is an example of interaction where the Alfv\'en vortex approximation is no longer valid \cite[]{sch94}. The main idea is that two aligned currents attract each other by the Lorentz force and can then coalesce. The generated vortex pattern has the characteristics of a collision of two vortices, and starts to deform the current distribution. Collisionless reconnection of the magnetic field takes place, changing the magnetic topology, and the magnetic flux is converted into electron momentum and ion vorticity, while the magnetic energy is transformed into electron energy \cite[]{kuv98,ber01}.

In the case of fast solar wind, we observed several examples of vortex chains where such interaction can take place, that could be related to a transient state of a collisionless reconnection. If it is the case, magnetic flux could be converted into electron momentum and ion vorticity, while the magnetic energy could be transformed into electron energy \cite[]{kuv98}. Unfortunately, due to very scarce particle measurements at time resolutions comparable with their kinetic scales, the heating process and the complicated phase-space interaction in turbulent solar wind still remain a puzzle. In particular, due to the low time resolution of the particle measurements on \textit{Cluster}, there are not enough points within the structures to study the heating and/or the energization of the particles. Moreover, sometimes the low resolution can generate unphysical effects due to the procedure of data sampling and averaging \cite[]{per14b}.

In March 2015, the Magnetospheric Multiscale (\textit{MMS}) mission, with four identically instrumented spacecraft as \textit{Cluster} but with a separation of $\sim 10$~km, has been launched. During Phase 2 of the mission, which started in February 2017, MMS apogee will be raised up to 25 $R_E$ and it will spend times in solar wind. In particular, after September 2017 the apogee will be located on the dayside and many intervals of solar wind data will be indeed gathered. In this context, particle distribution functions will be measured with high time resolution (30~ms for electron distributions and 150~ms for ions) and more detailed information will be obtained to better understand the problem of dissipation in a such collisionless plasma.

However, although \textit{MMS} has improved the temporal resolution for the particle measurements, angular/energy resolution still remains insufficient to completely resolve solar wind ions. A key insight in the study of turbulence, energy dissipation and particle energization in the near-Earth environment might be provided by the Turbulence Heating ObserveR (\textit{THOR}) mission \cite[]{vai16}, which is currently in the competitive study phase with two other missions at the ESA and it could be selected in end-2017. The main goal of this future space mission is to resolve kinetic scale processes, increasing angular and energy resolutions and the sensitivity of instruments, in particular for particle measurements.

\section*{acknowledgments}
We thank the FGM, STAFF, CIS, WHISPER, EFW instrument teams and the ESA Cluster Science Archive. Moreover we thank PEACE team at MSSL for the electron data. D. P. would like to acknowledge A. Mangeney, S. Toledo Redondo, B. Lavraud, D. Burgess and the joint ESTEC/ESAC heliophysics group for helpful discussions. O. A. and M. M. thank CNES for the instrumental development of {\it Cluster}/STAFF and for the financial support. Authors would like to acknowledge the anonymous Referee for his/her comments, which helped to improve the manuscript.  

\appendix
 
\subsection{Multi-point Signal Resonator technique}
\label{subsec:filt}

In the present paper we show that the solar wind turbulence is strong with the presence of coherent structures, characterized by coherence (constant phase) over many scales. However, in the context of the solar wind turbulence, many studies have been performed by using the {\it{$k$-filtering}} technique which was developed for the analysis of multi-point magnetometer data from the Cluster mission \cite[]{pin91}. The technique requires the assumptions of weak stationarity of the time series and that the signal can be described as a superposition of plane waves with random phases and a small component of isotropic noise. The Multi-point Signal Resonator (MSR) technique \citep[]{nar11a} is an extension of the {\it{$k$-filtering}} technique and requires the same assumptions. The main difference is that the MSR technique uses an additional filter based on the MUltiple SIgnal Classification (MUSIC) algorithm \cite[]{schm86}, to improve the signal to noise ratio of the power spectrum in wave-vector space $P(\omega_{sc},\mathbf{k})$. This method has also been validated for a synthetic signal which consists of random phase plane waves and non-random coherent structures \citep[]{rob14}. Moreover, \cite{rob14} showed that the presence of coherent components in the signal did not affect the recovery of any incoherent components. Using this approach the wave-number $k$ with the maximum power in the signal at a given spacecraft frequency $\omega_{sc}$ can be obtained without the need of Taylor's hypothesis. Moreover, the plasma frame frequency and the phase speed of the fluctuations can be obtained by Doppler shifting to the plasma frame according to the equations $\omega_{pla}=\omega_{sc}-\mathbf{k}\cdot\mathbf{v_{sw}}$ and $v_{ph}=\omega_{pla}/k$.

To verify the applicability of the MSR technique on a stream characterized by the presence of coherent structures, where its assumptions are seemingly in contradiction with the idea of strong turbulence, we perform the MSR analysis on this interval of fast solar wind, in two different ways combining data from the four spacecraft. The first is on the three components of the magnetic field which is dominated by incompressive fluctuations. The second method will focus only on the compressive fluctuations $\delta B_{\parallel}$ of the magnetic field by using a single input (the magnitude of the magnetic field) at each spacecraft. The application of the method to a single time series at each craft is discussed in detail by \cite{rob17}.

It is worth pointing out that this technique has some limitations. First of all, the fluctuations which can be surveyed are limited to scales comparable to the size of the Cluster tetrahedron. The maximum wave-number is given by the relation $k_{max}=\pi/d$ where $d$ is the mean spacecraft separation \cite[]{sah10,rob14}. Additionally the tetrahedron needs to be close to regular such that the spacecraft sample homogeneously in space. In this case planarity and elongation parameters \cite[]{Robert1998} are low $P\sim E \lesssim 0.15$, indicating that the geometry is close to that of a regular tetrahedron. Moreover, there are two sources of error for the plasma frame speed of the fluctuations. The first one consists in the error on determining the solar wind velocity which dominates the estimation of the plasma frame frequency $\omega_{pla}$, which is assumed to be $10\%$ \cite[]{mart93}. The second source of error is on the determination of the wave-number from the method. \cite{sah10} demonstrated that for a plane wave the wave-number is identified with a relative error of $10\%$ at a wave-number of $k_{max}/25$, that decreases quickly to $1\%$ at $k_{max}$. Furthermore, using the same approach when only a single time series is used at each spacecraft (as is the case when using the $|B|$ as an input) the errors in determining the wave-vector are similar \cite[]{rob17}.

\begin{figure}
\begin{center}
\includegraphics [width=8.8cm]{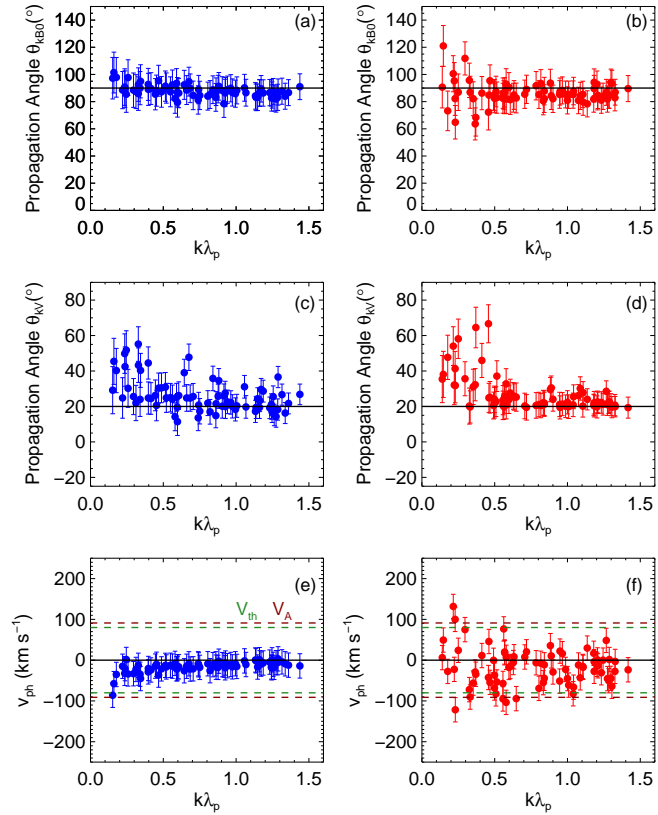}
\caption{Results from the $k$-filtering analysis for total (left panel, blue color) and compressive (right panel, red color) magnetic fluctuations. Panels (a) and (b): propagation angles with respect to the global magnetic field direction, $\theta_{kB_0}$. Horizontal black solid lines indicate $\theta_{kB_0}=90^{\circ}$. Panels (c) and (d): propagation angles with respect to the global velocity field direction. Horizontal black solid lines indicate $\theta_{kV}=20^{\circ}$. Panels (e) and (f): phase speeds of the fluctuations. Horizontal black solid lines indicate $V_{ph}=0$~km/s, while the Alfv\'en speed (horizontal dark purple dashed lines) and the proton thermal speed (horizontal green dashed lines) are given as references.}
\label{fg:kfil}
\end{center}
\end{figure}

The results of this analysis, for the considered interval of fast solar wind, are presented in Figure~\ref{fg:kfil}, where panels (a), (c) and (e) show the results when applied to the total magnetic fluctuations, while panels (b), (d) and (f) when it is applied to the fluctuations in the magnitude. It is worth pointing out that, unlike the timing method where each events is individually analyzed, the MSR technique has a global vision of the whole interval. This would not only include coherent structures but could also contain power from other sources such as incoherent plasma waves. At each frequency the wavevector corresponding to the most energetic fluctuations is recovered; therefore, it is possible that the MSR method gives results exactly for the coherent structures in the paper, but we cannot rule out contributions from other sources. Indeed, the MSR technique shows results similar to the timing method. Both total and compressive fluctuations are characterized by a small phase speed ($v_{ph} < V_{A}, V_{th}$) and propagation angles almost perpendicular to the global magnetic field ($\theta_{kB_{0}}\sim90^{\circ}$), with $\theta_{kV}\sim25^{\circ}$, even if for the compressive fluctuations a larger spread in both these values is observed. Nevertheless, the fact that $k$ is quasi-aligned with the solar wind speed could be an effect of the sampling direction. Therefore, it is not possible to conclude on the (non)gyrotropy of the turbulent fluctuations, because no information of the $k$ perpendicular to the solar wind velocity are known. A numerical study to test this point is needed and it will be the subject for a future work. 

The results of the MSR technique on the phase speed and propagation angle are consistent with previous case \cite[]{sah10,rob13} and statistical studies \cite[]{rob15a,Perschke2016} in the solar wind. This has variously been interpreted as evidence of quasi-linear waves that propagate slowly in the plasma frame (such as the quasi-perpendicular Kinetic Alfv\'en Wave \cite[]{sah10}) or coherent structures with $k_{\perp}\gg k_{\parallel}$ which are advected by the plasma bulk flow or a combination of these two phenomena \cite[]{rob13,rob15a}. However, in the present work, by looking directly at the magnetic fluctuations around ion scales in the region of high intermittency, it has been clearly shown that the considered time interval is covered by coherent structures (making up $\sim 30\%$ of the interval). These intermittent structures are characterized by a perpendicular wave-vector anisotropy (see the distribution of $\theta_{nB}$ in Figure~\ref{fg:direction}) and small velocities of propagation in the plasma frame (see Figure~\ref{fg:speed}). Generally structures which have very weak compressibility such as current sheets and vortices are advected with the flow. Meanwhile some compressive vortices are measured by the timing method to have higher speed and could account for the larger spread of values seen here.

\end{document}